\pdfoutput=1
\documentclass[twocolumn,showpacs,superscriptaddress,preprintnumbers,prb]{revtex4}

\usepackage{graphicx}
\usepackage{dcolumn}
\usepackage{bm}

\begin{document}
\title{Superconducting pairing and the pseudogap in nematic striped La$_{2-x}$Sr$_x$CuO$_4$
}


\author{S. Sugai}
\email{ssugai@pi.ac.ae}
\affiliation{Department of Physics, Arts and Science, Petroleum Institute, 
P. O. Box 2533, Abu Dhabi, UAE}
\affiliation{Department of Physics, Faculty of Science, Nagoya University, 
Furo-cho, Chikusa-ku, Nagoya 464-8602, Japan}
\author{Y. Takayanagi}
\affiliation{Department of Physics, Faculty of Science, Nagoya University, 
Furo-cho, Chikusa-ku, Nagoya 464-8602, Japan}
\author{N. Hayamizu}
\affiliation{Department of Physics, Faculty of Science, Nagoya University, 
Furo-cho, Chikusa-ku, Nagoya 464-8602, Japan}
\author{T. Muroi}
\affiliation{Department of Physics, Faculty of Science, Nagoya University, 
Furo-cho, Chikusa-ku, Nagoya 464-8602, Japan}
\author{R. Shiozaki}
\affiliation{Department of Physics, Faculty of Science, Nagoya University, 
Furo-cho, Chikusa-ku, Nagoya 464-8602, Japan}
\author{J. Nohara}
\affiliation{Department of Physics, Faculty of Science, Nagoya University, 
Furo-cho, Chikusa-ku, Nagoya 464-8602, Japan}
\author{K. Takenaka}
\affiliation{Department of Crystalline Materials Science, Nagoya University, 
Furo-cho, Chikusa-ku, Nagoya 464-8603, Japan}
\author{K. Okazaki}
\affiliation{Institute for Solid State Physics (ISSP), University of Tokyo,
Kashiwa, Chiba 277-8581, Japan
}

\date{\today}

\begin{abstract}
The individual $k||$ and $k\!\perp$ stripe excitations in fluctuating spin-charge 
stripes have not been observed yet.  
Raman scattering has a unique selection rule that the combination of two electric 
field directions of incident and scattered light determines the observed symmetry.  
If we set, for example, two electric fields to two possible stripe directions, 
we can observe the fluctuating stripe as if it is static.  
Using the different symmetry selection rule between the $B_{\rm 1g}$ two-magnon scattering 
and the $B_{\rm 1g}$ and $B_{\rm 2g}$ isotropic electronic scattering, 
we succeeded to obtain the $k||$ and 
$k\!\perp$ strip magnetic excitations separately in La$_{2-x}$Sr$_x$CuO$_4$.  
Only the $k\!\perp$ stripe excitations appear in the wide-energy isotropic electronic 
Raman scattering, indicating that the charge transfer is restricted to the direction 
perpendicular to the fluctuating stripe.  
This surprising restriction is reminiscent of the Burgers vector of an edge 
dislocation in metal.  
The edge dislocation easily slides perpendicularly to an inserted stripe and 
causes ductility in metal.  
Hence charges at the edge of a stripe move together with the edge 
dislocation perpendicularly to the stripe, while other charges are localized.  
A looped edge dislocation has lower energy than a single edge dislocation.  
The superconducting coherence length is close to the inter-charge stripe distance 
at $x\le 0.2$.  
Therefore we conclude that Cooper pairs are formed at looped edge dislocations.  
The restricted charge transfer direction naturally explains the opening of a pseudogap 
around $(0,\pi)$ for the stripe parallel to the $b$ axis and the reconstruction of the 
Fermi surface to have a flat plane near $(0,\pi)$.  
They break the four-fold rotational symmetry.  
Furthermore the systematic experiments revealed the carrier density dependence of the 
isotropic and anisotropic electronic excitations, the spin density wave and/or charge 
density wave gap near $(\pi/2,\pi/2)$, and the strong coupling between the electronic 
states near $(\pi/2,\pi/2)$ and the zone boundary phonons at $(\pi,\pi)$.  
\end{abstract} 

\maketitle
\section{Introduction}
Soon after the discovery of the high temperature superconductor, an 
incommensurate spin modulation was found by neutron 
scattering \cite{Yoshizawa,Birgeneau}.  
Stabilities of the superconducting states in spin stripes and spin 
plaquettes were investigated 
\cite{Machida,Zaanen1989,Kivelson,Zaanen,Kivelson2003,Sachdev,Zaanen2,Vojta}.  
The inhomogeneous structures were expected to solve the question that the 
two-dimensional Hubbard model may not stabilize the superconducting state 
\cite{Imada,Zhang,Aimi}.  
A periodic lattice modulation was found in La$_{2-x}$Sr$_x$CuO$_4$ (LSCO) by 
EXAFS \cite{Bianconi,Saini} 
and the atomic pair distribution function analysis of neutron 
diffraction \cite{Bozin}.  
Tranquada {\it et al}. found the spin-charge stripe structure in 
superconductivity suppressed La$_{1.48}$Nd$_{0.4}$Sr$_{0.12}$CuO$_4$ 
(LNSCO) by neutron scattering \cite{Tranquada}.  
Neutron scattering could not detect the charge density, so that the charge 
modulation was supposed from the lattice modulation.  
The charge modulation was certified by resonant soft X-ray scattering 
(RSXS) \cite{Abbamonte}.  
Yamada's group disclosed that the stripe structure in LSCO is 
ubiquitous in the doped insulating and superconducting phases, 
but it disappears outside of those phases 
\cite{Yamada,Wakimoto,Matsuda,Fujita2002,Christensen,Wakimoto2004,Wakimoto2007,
Matsuda2,Matsuda3}.  
The stripes in metal is fluctuating, because the incommensurate spots 
are observed in inelastic neutron scattering with a spin gap of about 5 
meV \cite{Lee2000}.  
When the fluctuation stops, the electronic state becomes insulating and 
superconductivity is suppressed in La$_{2-x}$Ba$_x$Cu$_2$O$_4$, LNSCO and 
Zn-doped LSCO with $x=1/8$\cite{Luke,Kumagai} as observed by neutron 
scattering \cite{Fujita} and $\mu$SR \cite{Watanabe,Nachumi,Adachi}.  

Fluctuation of the nematic stripe is important to induce a metallic 
conductivity \cite{Kivelson,Zaanen,Kivelson2003,Sachdev,Zaanen2,Vojta}.  
However, it is very difficult to observe the fluctuating stripe.  
The anisotropic magnetic excitations for $k||$ and $k\!\perp$ stripe 
have not been observed.  
The exception is the anisotropic low-energy excitations in the 
chain direction of YBa$_2$Cu$_3$O$_{7-\delta}$ (YBCO) \cite{Mook,Hinkov}.  
The high-energy magnetic excitations are presented by the so-called 
``hour-glass'' dispersion in the magnetic susceptibility versus wave vector 
\cite{Arai,Bourges,Hayden,Tranquada2,Hinkov,Christensen,Stock2005,Vignolle,
Hinkov2007,Kofu,Reznik,Lipscombe,Xu,Stock}.  
In the metallic state the four incommensurate scattering spots at 
$(\pi \pm \delta,\pi)$ and $(\pi,\pi \pm \delta)$ converge at the 
crossing point energy (resonance energy) as the energy increases and then diverge 
again in the directions rotated by $45^\circ$ from the low-energy dispersion directions.  
The magnetic excitations are interpreted by dynamical 
stripes \cite{Batista,Vojta2004,Uhrig,Seibold,Vojta2006,Seibold2} 
and interacting itinerant fermion liquid \cite{Morr,Eremin,Norman,Eremin2007}.  

Raman scattering has the unique selection rule that the combination of 
incident and scattered light polarizations determines the observed symmetry.  
If we choose, for example, the electric field of incident light to 
one of the possible stripe direction and the electric field of scattered 
light to the other possible stripe direction, we can observe the same Raman 
spectra without regard to the two possible stripe directions, because the Raman spectra 
are symmetric for the exchange of incident and scattered light.  
If the magnetic Raman scattering process is only one, we cannot separate 
the $k||$ and $k\!\perp$ stripe spectra.  
Fortunately two mechanisms with different symmetries contribute to high-energy 
magnetic Raman scattering.  
We can choose two different symmetries, $B_{\rm 1g}$ and $B_{\rm 2g}$.  
The $B_{\rm 1g}$ spectra are obtained in the polarization combination $(x,y)$ and 
the $B_{\rm 2g}$ spectra in the $(a,b)$, where $(x,y)$ denotes that incident light with 
the electric field parallel to the $x$ direction illuminates the sample and 
scattered light with the electric field parallel to the $y$ direction is measured.  
Here the tetragonal notation is used.  
$a$ and $b$ are the directions connecting Cu-O-Cu and $x$ and $y$ are the directions 
rotated by $45^{\circ}$.  
The two possible stripe directions are 
$x$ and $y$ in the insulating phase \cite{Wakimoto} and $a$ and $b$ in 
the metallic phase \cite{Yamada}.  
High-energy magnetic scattering is caused by two different mechanisms, 
two-magnon scattering \cite{Fleury,Parkinson,Canali} and electronic scattering 
\cite{Shastry,Shastry1991,Devereaux1994,Devereaux1995,Freericks2001,Freericks,
Shvaika2005,Devereaux,Medici}.  
Two-magnon scattering is active even in the insulating antiferromagnetic phase, 
while electronic scattering is caused by doped carriers.  
Utilizing this technique we succeeded to observe the individual $k||$ 
and $k\!\perp$ stripe magnetic excitations in fluctuating stripes.  

Many experimental results of Raman scattering were reported with respect to the 
high temperature superconductivity.  
The superconducting gaps in hole-doped superconductors were investigated by low-energy 
Raman scattering 
\cite{Chen1993,Blumberg,Chen1998,Liu,Naeini1999,Sugai2000,Opel,Hewitt,Gallais,Venturini,Masui,Sugai,Tacon2005,
Tassini2005,Tacon,Sugai2,Tacon2007,Tassini,Guyard,Sugai3,Masui2,Bakr,Blanc,Muschler,
Munnikes,Sugai4}.  
Two-magnon excitations and electronic excitations were investigated by wide-energy 
Raman scattering 
\cite{Lyons,Singh,Sulewski,SugaiTwoMag,Maksimov,Rubhausen,Blumberg,Liu,Naeini1999,Opel,Naeini,
Nachumi2002,Sugai,Machtoub,Tassini,Muschler,Caprara}.  
Two-magnon scattering is active only in the $B_{\rm 1g}$ spectra \cite{Lyons,Singh}.  
If electronic Raman scattering is treated without strong correlation, the spectral 
energy range is less than a few tens cm$^{-1}$ because of the momentum conservation 
with light.  
In calculation the $B_{\rm 1g}$ spectra is much stronger than the $B_{\rm 2g}$ spectra, 
because the $B_{\rm 1g}$ intensity is proportional to the square of the nearest 
neighbor hopping 
integral while the $B_{\rm 2g}$ intensity is proportional to the square of the diagonal 
next nearest neighbor hopping integral.  
Introduction of the strong correlation expands the spectral energy range to 1 eV 
through the self energy of the Green's function \cite{Shastry}.  
In the dynamical mean field theory the $k$ dependence of the self energy is 
ignored \cite{Georges,Freericks2001,Freericks,Shvaika2005,Devereaux,Medici}.  
The electron-radiation interaction Hamiltonian is expanded with respect to ${\bf A}$.  
The second order perturbation of the ${\bf A}$ linear term gives the resonant term 
in the scattering susceptibility and the first order perturbation of the ${\bf A}$ 
quadratic term gives the nonresonant susceptibility.  
Two-magnon scattering in insulator is given by the resonant term.  
Electronic scattering is composed of the nonresonant term and the resonant term.  
The intensity of the $B_{\rm 2g}$ channel mainly comes from the resonant term.  
The calculated $B_{\rm 2g}$ intensity is much smaller than the $B_{\rm 1g}$ 
intensity \cite{Shvaika2005}.  

The present experiment revealed that the $B_{\rm 2g}$ wide-energy spectra become the 
same as the $B_{\rm 1g}$ spectra above 2000 cm$^{-1}$ in the underdoped phase, if the 
two-magnon scattering is removed from the $B_{\rm 1g}$ spectra.  
It indicates that the dispersion becomes isotropic in $k$ space as the energy moves 
away from the chemical potential.  
It is also observed in the $A_{\rm 1g}$ spectra as an increasing screening effect at 
high energies.  
The common component decreases in the overdoped phase and 
the $B_{\rm 1g}$ spectra becomes stronger than the $B_{\rm 2g}$ spectra.  
It is approaching the calculated electronic states without stripe structure 
in electronic Raman scattering \cite{Shvaika2005}.  
We found a hump from 1000 to 3500 cm$^{-1}$ in the common spectra.  
The energy changes as carrier density increases and the intensity increases as 
temperature decreases.  
The hump can be well understood by the separated dispersion segments in the $k\!\perp$ 
stripe dispersion calculation by Seibold and Lorenzana \cite{Seibold,Seibold2}.  
The $k||$ stripe dispersion decreases in energy as well as 
the decrease in the high-energy spin susceptibility.  
On the other hand the $k\!\perp$ stripe dispersion is separated into $d$ 
segments without changing the overall dispersion energy, where $d$ is 
the multiplication factor of the spin stripe width with respect to the 
original magnetic unit cell.  
The width $d$ decreases as $d=1/2x$ with increasing the carrier density 
from $x\approx 0$ to $x=1/8$ and then keeps constant above $x=1/8$ \cite{Yamada}.  

From the analysis of the Raman spectra, we found that the electronic 
scattering spectra have only $k\!\perp$ stripe excitations.  
It means that the charge transfer is restricted only to the $k\!\perp$ 
stripe direction.  
This surprising result is reminiscent of the Burgers vector of an edge 
dislocation in metal \cite{Kleinert}.  
The edge dislocation and the screw dislocation easily slide and cause 
ductility in metal.  
In two-dimensional layer only the edge dislocation is available.  
The edge dislocation slides in the Burgers vector direction which is 
perpendicular to the inserted stripe.  
Charges at the edge of a stripe move together with the edge 
dislocation and other charges are localized, because 
$k||$ stripe excitation is not observed in the $B_{\rm 2g}$ electronic scattering.  
A looped edge dislocation connecting two charge stripes has lower energy 
than the single edge dislocation \cite{Zaanen}, because the spin alignments on both 
sides of the charge stripe have opposite phase \cite{Tranquada}.  
Zaanen \cite{Zaanen,Zaanen2} proposed a superconducting model generated 
by bosonized charges at edge dislocations.  
The spin-charge separation is not observed experimentally.  
Therefore it is supposed that Cooper pairs are formed at the moving edge 
dislocations.  
This model is supported by the experimental fact that the superconducting 
coherence length \cite{Wang,Wen} is close to the inter-charge stripe 
distance $d$ \cite{Yamada} at $x<0.2$.  
The coherence length is only twice of the inter-charge distance on the 
assumption that charges are uniformly distributed.  
The superconducting state is in the crossover regime between BCS 
(Bardeen-Cooper-Schrieffer) and BEC (Bose-Einstein condensation) \cite{Melo,Tsuchiya}.
The one-dimensional sliding motion of the charge can explain the pseudogap 
around $(0,\pi)$ in the underdoped phase.  
The $B_{\rm 2g}$ spectra have a low-energy hump composed of electron-phonon 
coupled states below 180 cm$^{-1}$.  
The spin density wave / charge density wave (SDW/CDW) gap and the superconducting 
gap appear in this sates.  

The electronic and two-magnon Raman scattering mechanisms 
are presented in Section~\ref{sec:theory}.  
The wide-energy Raman scattering, the analysis with respect to the anisotropy 
or isotropy in $k$ space, $k||$ and $k\!\perp$ stripe excitations, and 
the low-energy Raman scattering are presented in Section~\ref{sec:Raman}.  
The pairing at the looped edge dislocations 
is proposed in Section~\ref{sec:pairing}.  
The one-dimensional sliding motion is applied to the pseudogap in 
Section~\ref{sec:pseudogap}.
Discussions are given in Section~\ref{sec:discussions}.  
The conclusion is presented in Section~\ref{sec:conclusion}.

\section{\label{sec:theory}Electronic Raman scattering and two-magnon 
Raman scattering}
\subsection{\label{subsec:eleram}Electronic Raman scattering}
Electronic Raman scattering in simple metal is caused by the first order 
perturbation of the ${\bf A}^2$ term and the second order of the ${\bf p}\cdot {\bf A}$ 
term in the electron-radiation interaction term $({\bf P}-\frac{e}{c}{\bf A})^2$.  
The matrix element is given by \cite{Wolff,Platzman}
\begin{widetext}
\begin{equation}
M=e^{\alpha}_ie^{\beta}_s \frac{1}{m} \left[\delta_{\alpha \beta} 
+\frac{1}{m} \left(\sum_{b} \frac{\langle a,{\bf k}_f|P_{\beta}|b,{\bf k}_i+{\bf q}\rangle \langle b,
{\bf k}_i+{\bf q}|P_{\alpha}|a,{\bf k}_i\rangle}
{\epsilon_{{\rm a},{\bf k}_i}-\epsilon_{b,{\bf k}_i+{\bf q}}+\hbar \omega_i}+X\right) \right],
\label{eq:matrix}
\end{equation}
\end{widetext}
where $X$ is the term with the different time order, $m$ the free electron mass, 
$e^{\alpha}_i$ and $e^{\beta}_s$ polarization vectors of incident and 
scattered light, ${\alpha}$ and $\beta$ the Cartesian coordinates, 
$\omega_i$ and ${\bf q}$ the incident photon energy 
and wave vector, $a$ and $b$ are the initial and intermediate electronic states, 
and ${\bf k}_i$ and ${\bf k}_f$ are the initial and final wave vectors of the electron.  
In the low energy and long wavelength approximation of the incident light, 
Eq.~(\ref{eq:matrix}) is the same form as the ${\bf k}\cdot {\bf p}$ perturbation.  
Hence Eq.~(\ref{eq:matrix}) becomes 
\begin{eqnarray}
M\approx e^{\alpha}_ie^{\beta}_s \frac{\partial ^2\epsilon({\bf k})}{\partial k_{\alpha}
\partial k_{\beta}}\approx e^{\alpha}_ie^{\beta}_s \left(\frac{1}{m^*}\right)_{\alpha \beta}
\label{eq:mass},
\end{eqnarray}
where $(1/{\bf m}^*)$ is the effective inverse mass tensor.  
The energy range of the Raman spectra is limited to less than a few tens cm$^{-1}$ 
due to the momentum conservation with light. 
The scattering intensity goes to zero as the momentum shift $q$ goes to zero.  

The Raman intensity is proportional to \cite{Monien,Cardona,Devereaux1995}
\begin{eqnarray}
\langle |\hat{\bf e}_i \cdot \left( \frac{1}{{\bf m}^*} \right) \cdot \hat{\bf e}_s|^2
\rangle _{\rm F}-|\langle \hat{\bf e}_i \cdot \left( \frac{1}{{\bf m}^*} \right) \cdot 
\hat{\bf e}_s\rangle _{\rm F}|^2
\label{eq:screening},
\end{eqnarray}
where $\langle \ \rangle_{\rm F}$ represents an average over the Fermi surface.  
The second term represents the  screening of the $A_{\rm 1g}$ spectra by plasma excitations. 
The $A_{\rm 1g}$ intensity is completely screened, if the energy dispersion is parabolic in 
$k$ space.  
The screening ratio can be used how the electronic states are isotropic around the Fermi surface.  
The $B_{\rm 1g}$ and $B_{\rm 2g}$ spectra are not screened.  

In the strongly correlated electron system, the upper and lower Hubbard bands 
of the Cu $3d_{x^2-y^2}$ level are taken into account.  
In the Hubbard model coupled with light the creation and annihilation operators of 
an electron develop as \cite{Shastry,Shastry1991}
\begin{eqnarray}
c^{\dagger}_{i\sigma}c_{j\sigma} \rightarrow c^{\dagger}_{i\sigma}c_{j\sigma} \exp 
\left(-i\frac{e}{\hbar c}\int_{i}^{j}{\bf A}\cdot d{\bm \ell} \right).
\end{eqnarray}
The interaction Hamiltonian between the Hubbard electron and an electromagnetic wave 
is expand to the second order in $A$ \cite{Shastry,Shastry1991} 
\begin{eqnarray}
H_{\rm int}&=&-\left(\frac{e}{\hbar c}\right) \sum_{\bf k} {\bf j}({\bf k})\cdot {\bf A}
(-{\bf k})\nonumber\\
&+&\frac{e^2}{2\hbar ^2c^2}\sum_{{\bf k},{\bf k}'}{\bf A}(-{\bf k})\tau_{\alpha,\beta}
({\bf k}+{\bf k}'){\bf A}(-{\bf k}'),  
\end{eqnarray}
where, the current operator is 
\begin{eqnarray}
j_{\alpha}({\bf q})=\sum_{\bf k} \frac{\partial \epsilon ({\bf k})}{\partial k_{\alpha}} 
c_{\sigma}^{\dagger} ({\bf k}+{\bf q}/2)c_{\sigma} ({\bf k}-{\bf q}/2),
\end{eqnarray}
and the stress tensor is 
\begin{eqnarray}
\tau_{\alpha,\beta}({\bf q})=\sum_{\bf k} \frac{\partial ^2\epsilon({\bf k})}{\partial 
k_{\alpha}\partial k_{\beta}} c_{\sigma}^{\dagger}({\bf k}+{\bf q}/2) c_{\sigma}({\bf k}-{\bf q}/2).
\end{eqnarray}
The Raman matrix element of the nonresonant term is given by the first order perturbation 
of $\tau$ 
\begin{eqnarray}
\langle f|M_{\alpha,\beta}^{N}({\bf q})|i\rangle =\langle f|\tau_{\alpha,\beta}({\bf q})|i\rangle 
\end{eqnarray}
and the resonant term is given by the second order perturbation of $\bf j$ 
\begin{eqnarray}
\langle f|M_{\alpha,\beta}^{R}({\bf q})|i\rangle &=&\sum_{\nu} \left( \frac{\langle f|j_{\beta}
({\bf k}_f)|\nu\rangle \langle \nu |j_{\alpha}(-{\bf k}_i)|i\rangle }
{\epsilon_{\nu}-\epsilon_i-\hbar \omega_i}\right. \nonumber\\
&+&\left. \frac{\langle f|j_{\alpha}(-{\bf k}_i)|\nu \rangle \langle \nu|j_{\beta}
({\bf k}_f)|i\rangle }{\epsilon_{\nu}-\epsilon_i+\hbar \omega_f} \right). 
\label{eq:resonant}
\end{eqnarray} 
An electron transferred to the neighboring site is excited to the upper Hubbard band in the 
intermediate state.  
The charge transfer excitation energy with double occupancy is close to the incident photon 
energy, so that the scattering is resonantly enhanced.  

In the dynamical mean field theory the imaginary part of the Raman 
susceptibility of the nonresonant term is given by \cite{Freericks2001,Freericks,Devereaux}
\begin{eqnarray}
\chi''(\omega)&=&\int d\epsilon V(\epsilon)\int d\omega' A(\epsilon,\omega')\nonumber\\
& &\times A(\epsilon,\omega'+\omega)[f(\omega')-f(\omega'+\omega)]
\label{eq:dmfsus},
\end{eqnarray}
where the form factor $V(\epsilon)$ is 
\begin{eqnarray}
V(\epsilon)=\sum_{\bf k} \biggl[ \frac{\partial ^2\epsilon(\bf k)}
{\partial k_i\partial k_j} \biggr] ^2 \delta[\epsilon-\epsilon(\bf k)],
\label{eq:v}
\end{eqnarray}
where $ij=xy$ for $B_{\rm 1g}$ and $ij=ab$ for $B_{\rm 2g}$.  
$f(\omega)$ is the Fermi-Dirac distribution function.  

The one-particle spectral function $A$ is the imaginary part of the Green function 
\begin{eqnarray}
A(\epsilon,\omega)=-\frac{1}{\pi} {\rm Im} \frac{1}{\omega+\mu-\Sigma(\omega)-\epsilon},
\label{eq:spfn}
\end{eqnarray}
where $\Sigma$ is the self energy representing the interactions with other particles.  
$\Sigma$ is independent of $\bf k$ in the dynamical mean field theory.  
The spectral function is composed of a coherent peak (quasi-particle peak) and two 
incoherent parts.  
The scattering intensity from the coherent peak goes to zero as $q$ goes to zero, 
while the incoherent parts keep the intensity.

\begin{figure}
\begin{center}
\includegraphics[trim=0mm 0mm 0mm 0mm, width=8cm]{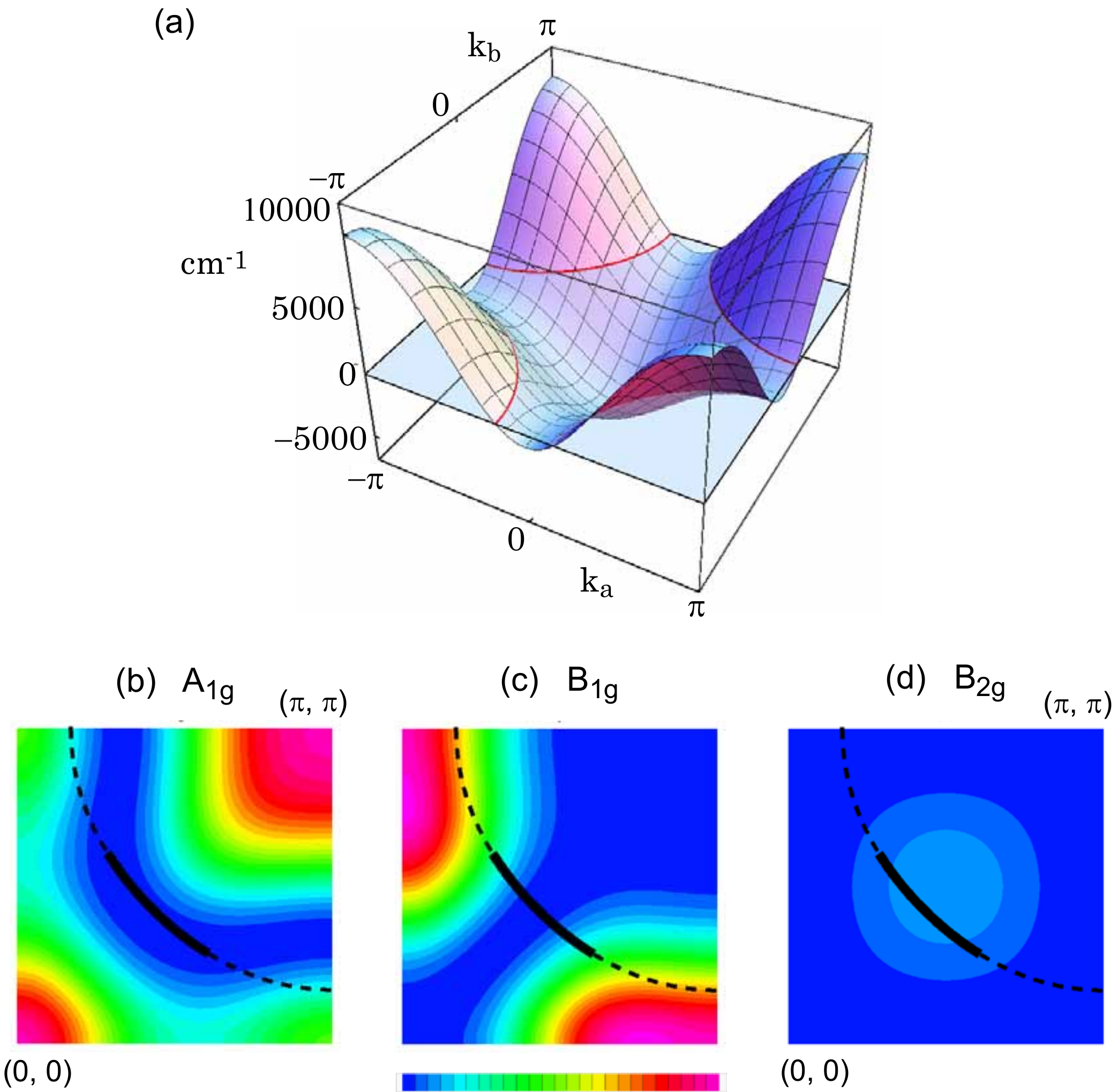}
\caption{(color online) 
(a) The electron energy dispersion in La$_{1.93}$Sr$_{0.07}$CuO$_4$ using the parameters 
reported by ARPES \cite{Yoshida}.  
The contour maps of $|\partial ^2\epsilon({\bf k})/\partial k_{\alpha} \partial_{\beta}|^2$ 
for (b) $A_{\rm 1g}$, (c) $B_{\rm 1g}$, and (d) $B_{\rm 2g}$.  
The solid line shows the Fermi arc and the dashed line shows the depleted Fermi surface 
(Pseudogap).  
}
\end{center}
\end{figure}
Figure 1(a) shows the electron energy dispersion of the tight binding model 
\begin{eqnarray}
\epsilon(k_a, k_b)&=&\epsilon_0-2t(\cos k_a+\cos k_b)-4t'\cos k_a \cos k_b\nonumber\\
& &-2t''(\cos 2k_a+\cos 2k_b),
\label{eq:tightbinding}
\end{eqnarray}
where $t$, $t'$, and $t''$ are the first-, second-, and third-nearest neighbor 
hopping integrals between Cu sites.  
The parameters are $t=0.25$ eV, $t'=-0.17t$ ($-0.15t$, $-0.12t$), $t''=-t'/2$, and 
$\epsilon_0=0.55t$ ($0.81t$, $0.99t$) for $x=0.07$ (0.15, 0.3) by angle-resolved 
photoemission spectroscopy (ARPES) \cite{Yoshida}.  
The light blue plane in Fig. 1(a) shows the chemical potential $\mu=0$.  
Figure 1(b), (c), and (d) show $[(\partial ^2\epsilon/\partial k_x \partial k_x)^2+
(\partial ^2\epsilon/\partial k_y \partial k_y)^2+(\partial ^2\epsilon/\partial k_a 
\partial k_a)^2+(\partial ^2\epsilon/\partial k_b \partial k_b)^2]/4-[(\partial ^2
\epsilon/\partial k_x \partial k_y)^2+(\partial ^2\epsilon/\partial k_a \partial k_b)^2]/2$ 
for $A_{\rm 1g}$, $(\partial ^2\epsilon/\partial k_x \partial k_y)^2$ for $B_{\rm 1g}$ 
and $(\partial ^2\epsilon/\partial k_a \partial k_b)^2$ for $B_{\rm 2g}$ at $x=0.07$.  
The $B_{\rm 1g}$ intensity is given by 
$[t({\rm cos}k_a-{\rm cos}k_b)+4t''({\rm cos}2k_a-{\rm cos}2k_b)]^2$ and 
the $B_{\rm 2g}$ intensity by $[4t'{\rm sin}k_a\ {\rm sin}k_b]^2$.  
The $B_{\rm 1g}$ spectra observe near $(\pi,0)$ and the $B_{\rm 2g}$ spectra observe near 
$(\pi/2,\pi/2)$ \cite{Devereaux1994,Devereaux1995}.  
The intensity near $(\pi,0)$ and $(0,\pi)$ in $B_{\rm 1g}$ is much larger than that 
near $(\pi/2,\pi/2)$ in $B_{\rm 2g}$, because $t$ is much larger than $t'$.  
The Fermi surface at $x=0.07$ is shown by the thick line and the dashed line.  
The dashed line indicates the pseudogap formed in the underdoped phase \cite{Yoshida}.  

\begin{figure}
\begin{center}
\includegraphics[trim=0mm 0mm 0mm 0mm, width=7cm]{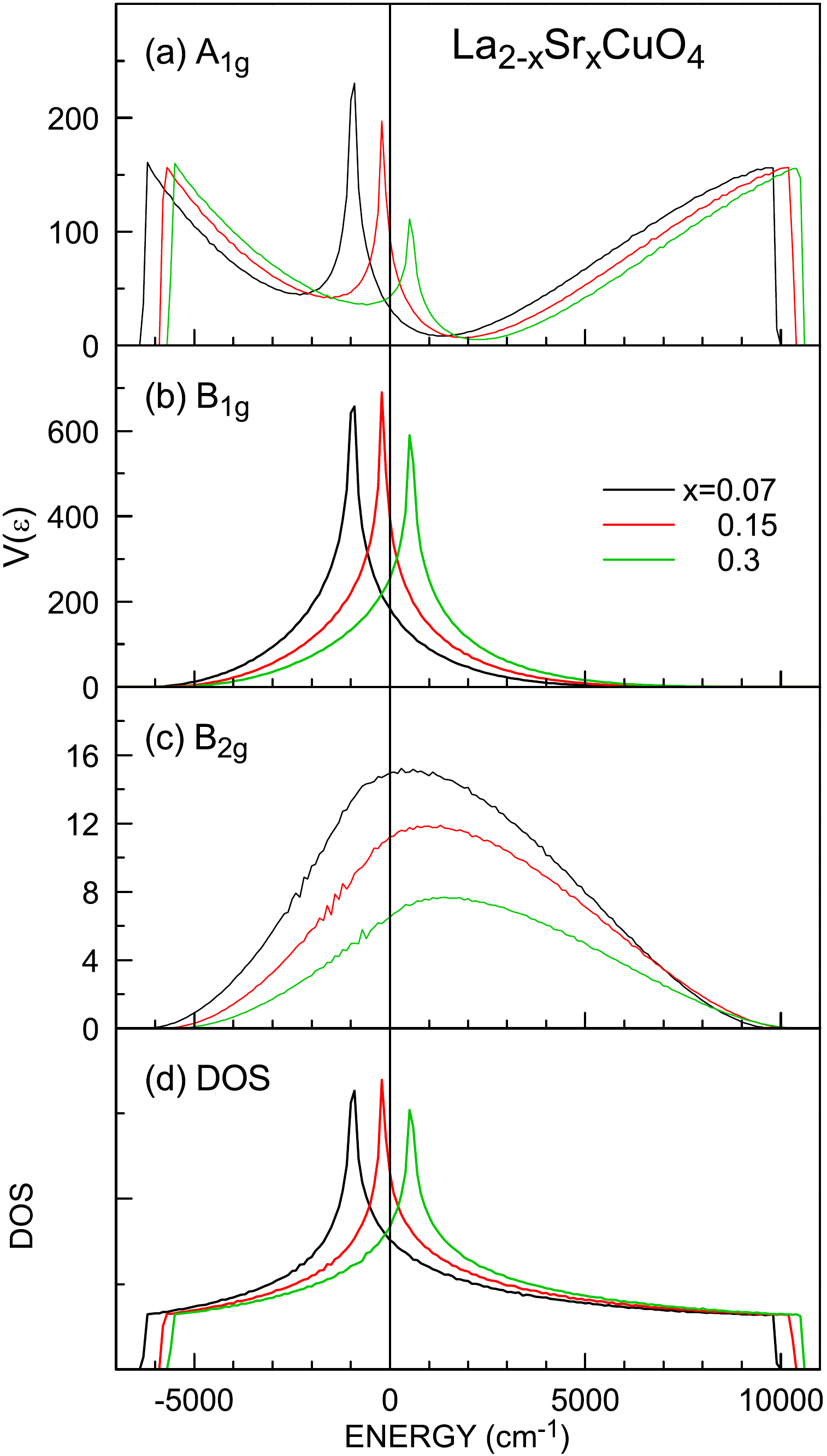}
\caption{(color online) 
The form factor $V(\epsilon)$ for (a) $A_{\rm 1g}$, (b) $B_{\rm 1g}$, and (c) $B_{\rm 2g}$.  
(d) The density of states.  
The parameters for the electron dispersion were given by ARPES \cite{Yoshida}.  
}
\end{center}
\end{figure}
In the dynamical mean field theory the difference between the $B_{\rm 1g}$ and 
$B_{\rm 2g}$ spectra comes from the $V(\epsilon)$ in Eq.~(\ref{eq:v}).  
Figure 2 shows the $V(\epsilon)$ for (a) the $A_{\rm 1g}$, (b) $B_{\rm 1g}$ and 
(c) $B_{\rm 2g}$ symmetries, and (d) the density of states.  
The chemical potential is energy zero.  
The intensity of the $A_{\rm 1g}$ spectra increase as energy shift increases, while 
those of the $B_{\rm 1g}$ and $B_{\rm 2g}$ spectra decrease at high energies.  
The present experiment, however, revealed that the intensity of the $A_{\rm 1g}$ 
spectra more rapidly decreases than the $B_{\rm 1g}$ and $B_{\rm 2g}$ spectra at 
high energies, indicating that the screening effect increases at high energies.  
The peak positions in $A_{\rm 1g}$ and $B_{\rm 1g}$ shift from $\epsilon<0$ 
to $\epsilon>0$ in Fig. 2, because the zone boundary point of the 
Fermi surface changes from $(0, \pi)-(\pi, \pi)$ to $(0, 0)-(0, \pi)$.  
The intensity of the $B_{\rm 2g}$ top is about 1/40 times of the $B_{\rm 1g}$ peak.  
The $B_{\rm 2g}$ intensity mainly comes from the resonant term, but the 
resonant scattering intensity is still much smaller than the $B_{\rm 1g}$ 
channel \cite{Shvaika2005}.  
The total intensity of the $B_{\rm 2g}$ channel is one order smaller than the 
$A_{\rm 1g}$ and $B_{\rm 1g}$ channels.
However, the present experiment revealed that the $B_{\rm 2g}$ intensity is the 
same order as the $B_{\rm 1g}$ intensity in the under doped phase.  

\begin{figure}
\begin{center}
\includegraphics[trim=0mm 0mm 0mm 0mm, width=7cm]{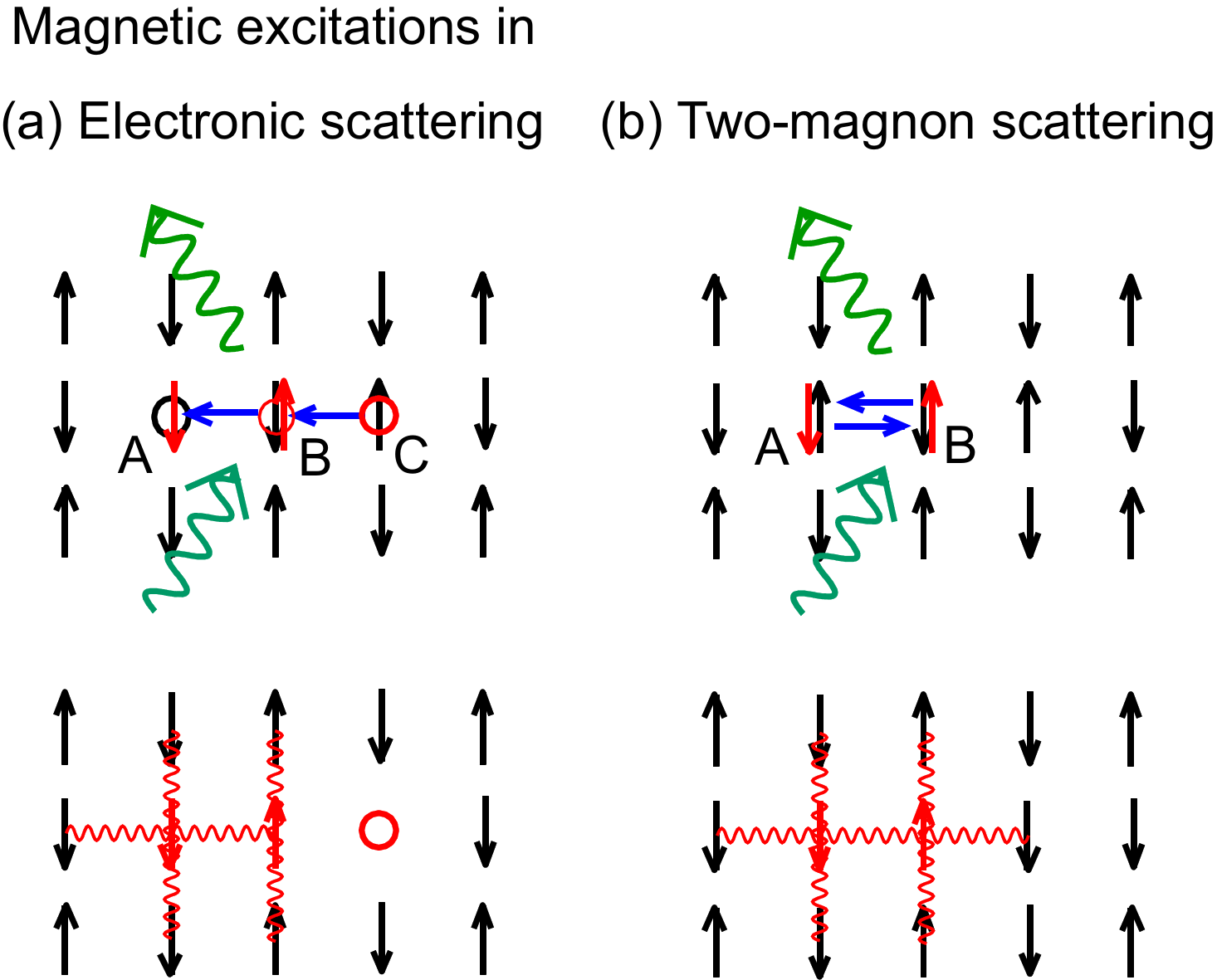}
\caption{(color online) 
The magnetic scattering processes induced by (a) a hole hopping from A to C 
in electronic scattering and (b) an exchange hopping in two-magnon scattering.  
The lower panels show final spin orientations.  
The bonds with increased exchange interaction energies are shown by the wavy lines.  
The spin excited states propagate as magnons.  
In case (a) two magnons are successively excited, while in case (b) two magnons are 
simultaneously excited.  
}
\end{center}
\end{figure}
Electronic Raman scattering detects magnetic excitations through the self-energy 
$\Sigma$ in Eq.~(\ref{eq:spfn}).  
A hole hopping from site A to the nearest neighbor site B is the same as 
a back hopping of an electron from B to A in Fig. 3(a).  
The coming electron spin is opposite to the stable spin direction at site A.  
Thus hole hopping causes the overturned spin trace shown in the lower panel.  
The red wavy lines show the increased energy bonds.  
The overturned spin excitation propagates as a magnon at each hopping from A to B 
and from B to C.

\subsection{\label{subsec:twomag}Two-magnon scattering}
Two-magnon scattering in the insulating phase is caused by the resonant term of 
Eq.~(\ref{eq:resonant}).  
A hole at A hops to the neighboring site B by absorbing light and the original 
hole at B hops to A by emitting light in Fig. 3(b) \cite{Shastry,Shastry1991}.  
This process gives the same interaction Hamiltonian as the Fleury-Loudon 
type \cite{Fleury,Parkinson}
\begin{eqnarray}
H_{\rm two-mag}=\sum_{kl}A({\bf e}_{i}\cdot {\bf r}_{kl})({\bf e}_{s}\cdot {\bf r}_{kl})
({\bf S}_k\cdot {\bf S}_l),
\label{eq:2mcb}
\end{eqnarray}
where ${\bf r}_{kl}$ is the unit vector connecting the $k$ and $l$ sites.  
Two-magnon scattering is active in $(aa)$ and $(xy)$ and inactive in $(ab)$.  
In $(xx)$ the two-magnon scattering Hamiltonian is the same as the system 
Hamiltonian 
\begin{eqnarray}
H=J\sum_{kl}{\bf S}_k\cdot {\bf S}_l.
\end{eqnarray}
except for the proportionality constant.  
Therefore two-magnon scattering is inactive, because the two-magnon Hamiltonian 
commutes to the system Hamiltonian.  
Two magnons are simultaneously excited, so that the two magnons interfere 
and the total energy is reduced from the independently excited two magnons by the 
magnon-magnon interaction energy which is close to the exchange interaction 
energy $J$ \cite{Fleury,Parkinson,Canali}.  
In the electronic scattering process in Fig. 3(a) the magnon excitation energy is 
included in the self-energy.  
A magnon is excited at each hopping process, so that the magnon-magnon interaction 
does not arise in the lowest order.  
The symmetry dependence of the magnetic Raman scattering mechanism is summarized 
in Table~\ref{tab:table1}.  
\begin{table}
\caption{\label{tab:table1}Symmetry dependence of the magnetic Raman scattering 
mechanism and the experimental results.  
}
\begin{ruledtabular}
\begin{tabular}{c | c c}
Spectral symmetry&$B_{\rm 1g}$&$B_{\rm 2g}$\\
\hline
Two-magnon scattering&Yes&No\\
Electronic scattering&Yes&Yes\\
\hline
Experimental results&$k||$+$k\!\perp$stripe&\ $k\!\perp$stripe \ \ \ \\
\end{tabular}
\end{ruledtabular}
\end{table}

\section{\label{sec:Raman}Raman scattering experiments}
\subsection{\label{subsec:exp}Experimental procedure}
In order to obtain the wide-energy spectra, the fine adjustment of the 
spectrometer is necessary.  
We used a triple-grating spectrometer with the same focusing lengths of 600 mm.  
The first two gratings are used as a filter to cut the direct laser light and the third 
grating is used to disperse the spectra.  
A Raman system is usually adjusted to measure molecular vibrations of less than 3000 
cm$^{-1}$, so that the measurement of large energy shift to 7000 cm$^{-1}$ is not warranted.   
The focusing point on the slit of the third spectrometer moves, as the central wavenumber 
of the spectrometer is driven into the infrared region, if the 
adjustment of the spectrometer is insufficient.  
It causes a decrease or increase of the intensity at high energy shift.  
We carefully adjusted the spectrometer every $3\sim 4$ months.  

Single crystals were synthesized by a traveling-solvent floating-zone method.  
The solvent were melted by the radiation from four halogen lamps with four elliptic mirrors.  
The excess oxygen in La$_2$CuO$_{4+\delta}$ crystals were reduced, but some excess 
oxygen remained.  
The oxygen is deficient in as-grown crystals of $x=0.2$ and 0.25.  
They were annealed in one atm oxygen gas at 600$^{\circ}$ 
for 7 days.  
Raman spectra were obtained on fresh cleaved single crystal surfaces in a 
quasi-back scattering configuration using 514.5 nm laser light.  
The incident angle from the normal direction of the sample surface was 30$^{\circ}$.  
The incident polarization direction was fixed to the horizontal direction ($p$-wave).  
The vertical or horizontal polarization of scattered light was selected.  
The $B_{\rm 1g}$ and $B_{\rm 2g}$ spectra were obtained by rotating the sample 
keeping other optical geometries in the same positions.  
The $B_{\rm 1g}$ and $B_{\rm 2g}$ spectra were obtained in the $(x,y)$ and $(a,b)$ 
polarizations, respectively. 
The $A_{\rm 1g}$ spectra were obtained from the calculation of the spectra 
$[(x,x)+(a,a)-(x,y)-(a,b)]/2$.  
The details of samples and Raman scattering were presented in our previous 
paper \cite{Sugai}.  
The wave number and polarization dependences of the optical system were carefully 
corrected using reflected light from a standard white reflection plate.  
The light source is a incandescent lamp with a known black body radiation temperature.  
The optical path for the measurement of the spectral efficiency was carefully adjusted 
to coincide with the Raman scattering experiment.  
The Raman intensity is proportional to 
$1/[\alpha_i(\omega_i)+\alpha_s(\omega_s)]$, where $\alpha$ 
is the absorption coefficient.  
The absorption coefficient of the incident laser light decreases by 0.7 times as 
the hole density increases from $x=0$ to 0.25, while it increases by 5 times at the 
energy shift of 7000 cm$^{-1}$.  
Therefore the absorption correction is necessary to compare the carrier density 
dependence.  
The absorption coefficient was obtained from far-infrared, visible and ultraviolet 
reflection spectroscopy by means of the Kramers-Kronig transformation.  
The details of infrared spectroscopy was presented in our previous paper \cite{Takenaka}

\subsection{\label{subsec:high}Wide-energy spectra : Anisotropic or isotropic 
electronic dispersion in $k$ space}
\begin{figure*}
\begin{center}
\includegraphics[trim=0mm 0mm 0mm 0mm, width=15cm]{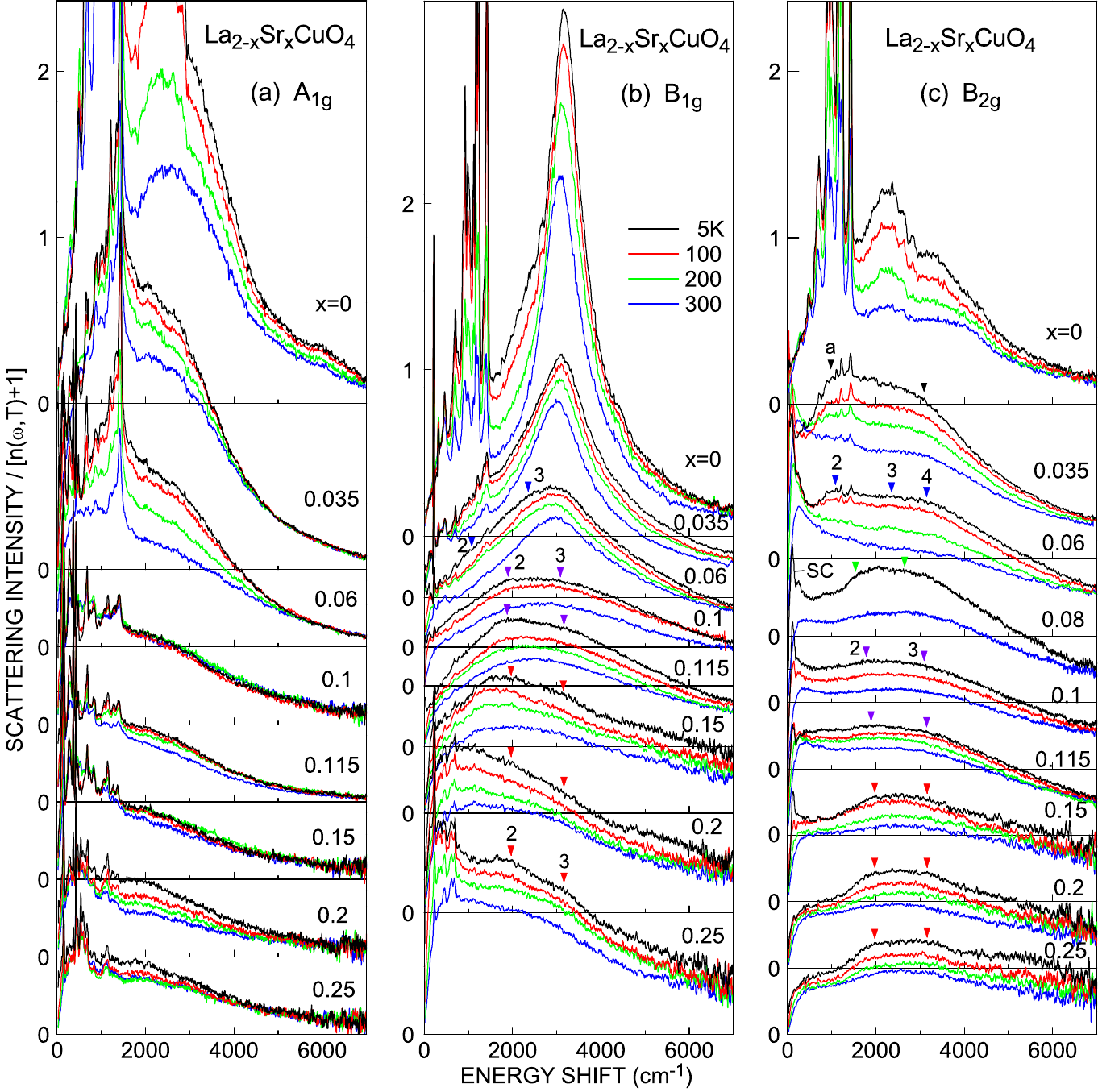}
\caption{(color online) 
Wide-energy (a) $A_{\rm 1g}$, (b) $B_{\rm 1g}$, and (c) $B_{\rm 2g}$ Raman spectra.  
The downward triangles correspond to the dispersion segments with the same number 
and color (blue, green and red) in the $k\!\perp$ stripe magnetic excitations in Fig. 11.
}
\end{center}
\end{figure*}
\begin{figure}
\begin{center}
\includegraphics[trim=0mm 0mm 0mm 0mm, width=8cm]{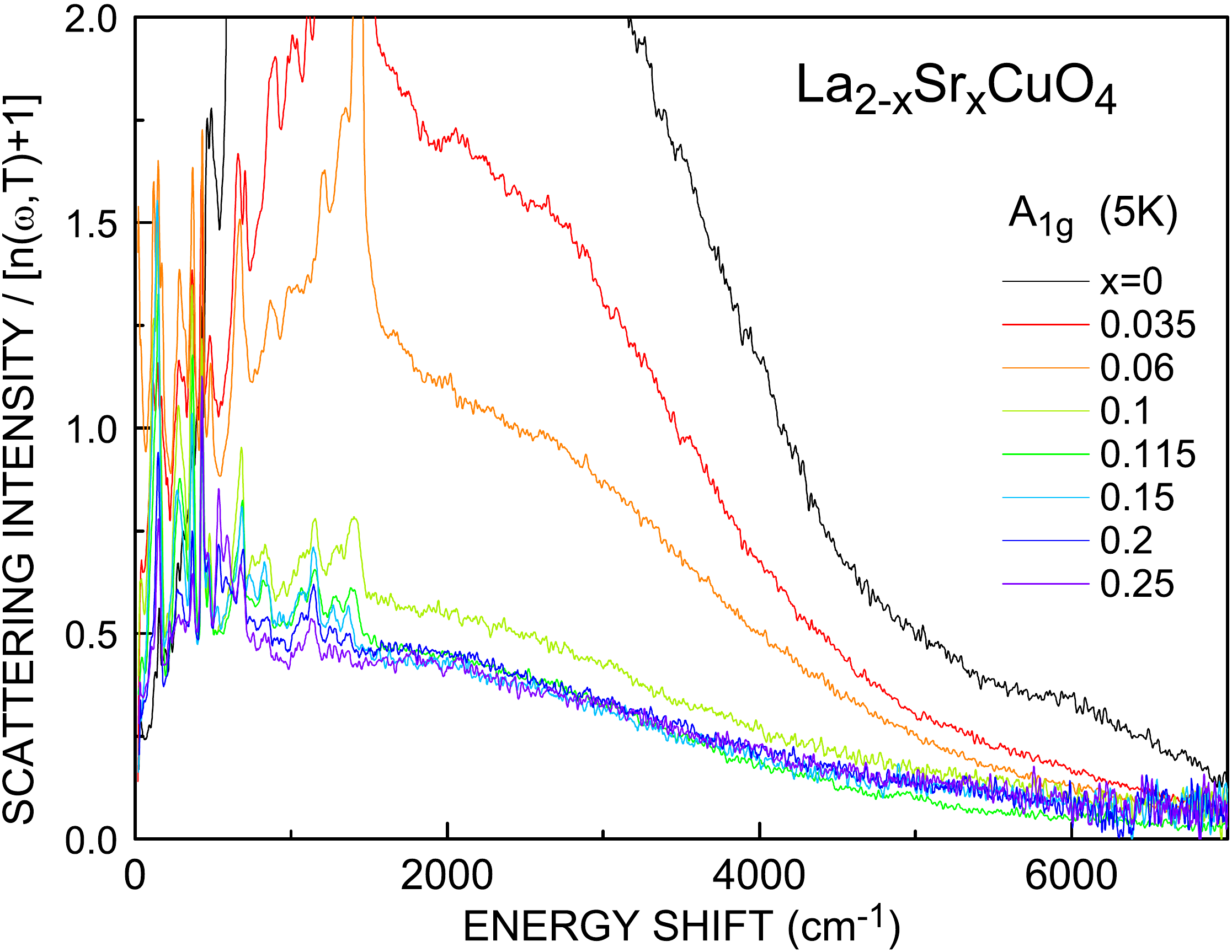}
\caption{(color online) 
Carrier density dependent wide-energy $A_{\rm 1g}$ spectra at 5 K.  
}
\end{center}
\end{figure}
\begin{figure}
\begin{center}
\includegraphics[trim=0mm 0mm 0mm 0mm, width=6cm]{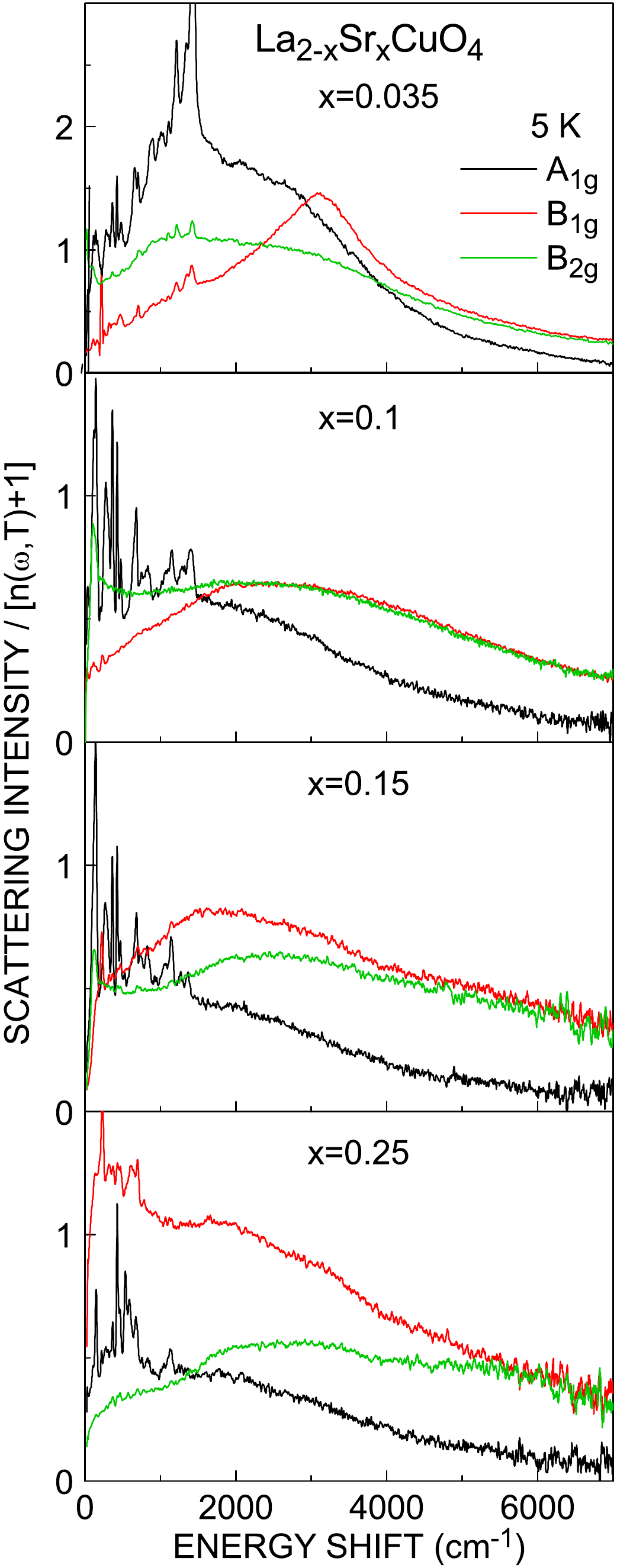}
\caption{(color online) 
Comparison among the $A_{\rm 1g}$, $B_{\rm 1g}$, and $B_{\rm 2g}$ spectra at 5 K.  
}
\end{center}
\end{figure}
Figure 4 shows the wide-energy Raman spectra.  
All the spectra are plotted in the same intensity scale.  
The sharp peaks from 700 to 1400 cm$^{-1}$ at $x = 0$ are two-phonon peaks.  
Four- and six-phonon peaks are observed in the $A_{\rm 1g}$ and $B_{\rm 2g}$ spectra.  
The multi-phonon spectra are 20 times stronger in the $A_{\rm 1g}$ spectra than in the 
$B_{\rm 1g}$ or $B_{\rm 2g}$ spectra at $x=0$.  
The multi-phonon intensity rapidly decreases to 1/60 at $x = 0.035$ and almost completely 
disappears at $x\ge 0.08$ in the $B_{\rm 1g}$ 
and $B_{\rm 2g}$ spectra, while the small intensity remains in the whole carrier 
density range in the $A_{\rm 1g}$ spectra.  
The 3170 cm$^{-1}$ peak in the $B_{\rm 1g}$ spectra at $x=0$ is the two-magnon peak.  
The 4400 cm$^{-1}$ subpeak at $4J$ appears in a polished sample, but almost 
completely disappears in a cleaved sample.  
The high-energy spectra are rather different from other groups \cite{Muschler,Caprara}.  
The difference comes from whether the crystal surface 
is cleaved or polished and how the spectral efficiency of the optical system 
is corrected.  

The wide-energy spectra are very different from the spectra expected from the form factor 
$V(\epsilon)$ in Fig. 2 with respect to the following points.  
(1) The $A_{\rm 1g}$ spectra decrease rapidly to high energy, which is contrary to the 
spectra expected from Fig. 2(a).  
(2) The $B_{\rm 2g}$ spectra have almost the same intensity as the $B_{\rm 1g}$ spectra 
in spite of very weak calculated intensity \cite{Shvaika2005}.  
The large difference between the experiment and the theory is caused by 
the deviation of the electronic states from the tight binding model of 
Eq.~(\ref{eq:tightbinding}).

Figure 5 shows the carrier density dependence of the $A_{\rm 1g}$ spectra at 5 K.  
The intensity rapidly decreases as the carrier density increases from $x=0$ to 0.1 
and then the spectra keep the same shape at $x\ge 0.1$.  
The spectra have a broad peak at 500 cm$^{-1}$ and a long tail to high energy at $x>0.1$.  
Figure 6 shows the comparison of the $A_{\rm 1g}$, $B_{\rm 1g}$, and $B_{\rm 2g}$ 
spectra at 5 K.  
The $B_{\rm 1g}$ and $B_{\rm 2g}$ spectra approach each other as the energy shift 
increases and become the same above 4000 cm$^{-1}$ at $x=0.035$, 2000 cm$^{-1}$ at $x=0.1$, 
4000 cm$^{-1}$ at $x=0.15$, and 5000 cm$^{-1}$ at $x=0.25$.  
It indicates that the anisotropy of the electron energy dispersion in $k$ space decreases 
as the energy moves away from the chemical potential, that is, the energy dispersion 
becomes isotropic at high energy shift.  
It is supposed that the unscreened $A_{\rm 1g}$ spectra also becomes the 
same as the $B_{\rm 1g}$ and $B_{\rm 2g}$ spectra at high energies.  
However, the $A_{\rm 1g}$ spectra are screened from Eq.~(\ref{eq:screening}), as the 
isotropy increases at high energies.  
As a result the $A_{\rm 1g}$ spectra are strongly depressed at high energies.  

\begin{figure}
\begin{center}
\includegraphics[trim=0mm 0mm 0mm 0mm, width=6cm]{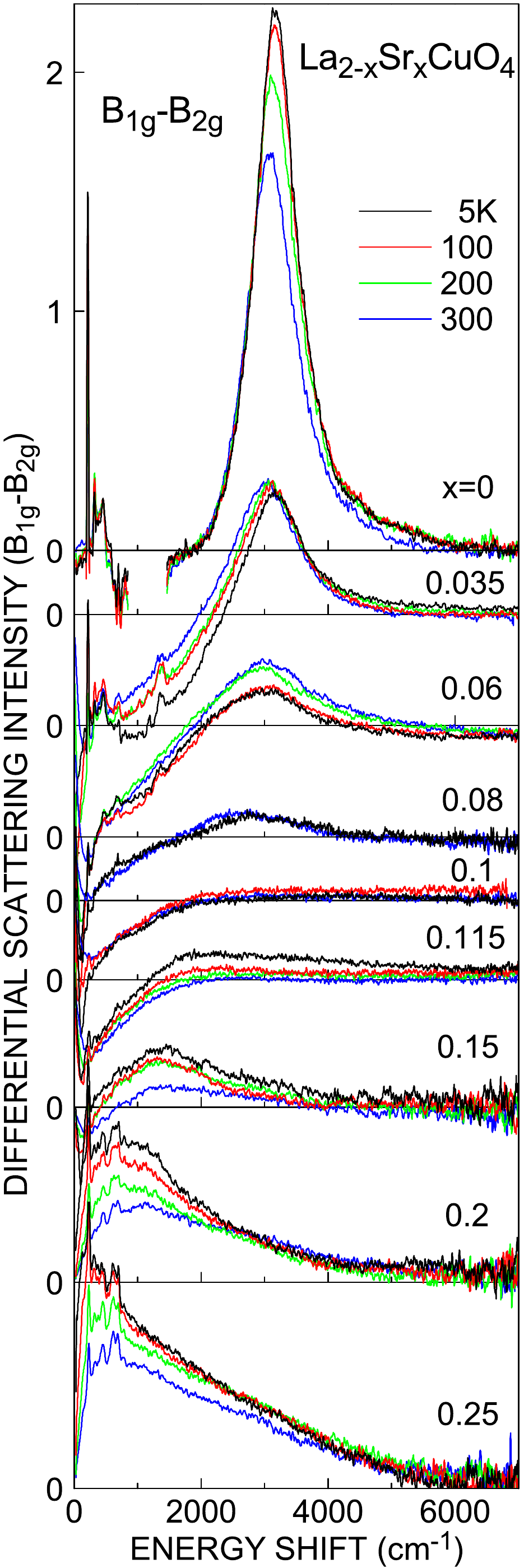}
\caption{(color online) 
Differential spectra between the $B_{\rm 1g}$ and $B_{\rm 2g}$ symmetries.  
}
\end{center}
\end{figure}
Figure 7 shows the differential spectra between the $B_{\rm 1g}$ and $B_{\rm 2g}$ 
symmetries.  
The two-magnon peak at $x=0$ is rather sharp, because the multi-phonon and 
electronic scattering components are removed.  
Two-magnon scattering is basically inactive in the $B_{\rm 2g}$ channel.  
As for the origin of the $B_{\rm 2g}$ two-magnon scattering, diagonal spin-pair 
excitations \cite{Singh} or the chiral spin excitations 
$\sum {\bf s}_i\cdot ({\bf s}_j \times {\bf s}_k)$ are proposed \cite{Shastry,Shastry1991}.  
The $B_{\rm 2g}$ two-magnon scattering is also canceled in Fig. 7.  
At $x=0.1$ the intensity above 2000 cm$^{-1}$ is zero, that is, the $B_{\rm 1g}$ 
and $B_{\rm 2g}$ spectra are the same.  
The $B_{\rm 1g}$ intensity decreases below 2000 cm$^{-1}$ due to the formation of 
the pseudogap around $(\pi,0)$.  
The similar structure is observed from $x=0.035$ to 0.115, if the two-magnon peak 
is removed.  
At $x=0.115$ the differential spectra are the same as $x=0.1$ from 300 K to 100 K.  
At 5 K a weak hump at 2010 cm$^{-1}$ and a long high-energy tail emerges.  
The hump enlarges and the peak energy softens, as the carrier density increases 
in the overdoped phase.  
The peak has a long tail to high energy.  
The intensity of the $B_{\rm 1g}$ spectra at $x=0.25$ is 4.1 times the $B_{\rm 2g}$ 
spectra at 150 cm$^{-1}$ and 1.8 times for the integrated intensity from 16 cm$^{-1}$ to 
6000 cm$^{-1}$.  
The two-magnon peak decreases in intensity and energy as $x$ increases from $x=0$ to 0.08.  
The two-magnon peak energy at $x\le 0.08$ and the hump energy at $x\ge 0.115$ are 
continued, although it is not clear whether the hump in the overdoped phase is related to 
the two-magnon scattering or not.  
The decreasing peak energy with increasing carrier density in the overdoped phase 
looks like the $B_{\rm 1g}$ spectra in the dynamical mean field calculation of the 
nonresonant term \cite{Freericks}.  
The characteristics hump at $1000- 3500$ cm$^{-1}$ in the $B_{\rm 2g}$ spectra of 
Fig. 4(c) is an important structure to assign the stripe excitations.  
The hump is enhanced as temperature decreases.  
The hump does not appear in the differential spectra of Fig. 7, representing 
that the $B_{\rm 1g}$ spectra have the same hump as the $B_{\rm 2g}$ spectra 
at all temperatures.  

The results of the differential spectra are summarized.  
In the underdopd phase (1) the electronic scattering spectra are same in the 
$B_{\rm 1g}$ and $B_{\rm 2g}$ channels above 2000 cm$^{-1}$, 
(2) the $B_{\rm 1g}$ intensity decreases below 2000 cm$^{-1}$, and 
(3) the two-magnon peak in the differential spectra decreases in intensity and 
energy, as the carrier density increases from $x=0$ to 0.08.  
In the overdoped phase (4) the $B_{\rm 1g}$ spectra get larger than the $B_{\rm 2g}$ 
spectra.  
In whole carrier density range (5) a hump appears at $1000-3500$ cm$^{-1}$ 
in both $B_{\rm 1g}$ and $B_{\rm 2g}$ spectra, as temperature decreases.

\begin{figure}
\begin{center}
\includegraphics[trim=0mm 0mm 0mm 0mm, width=8cm]{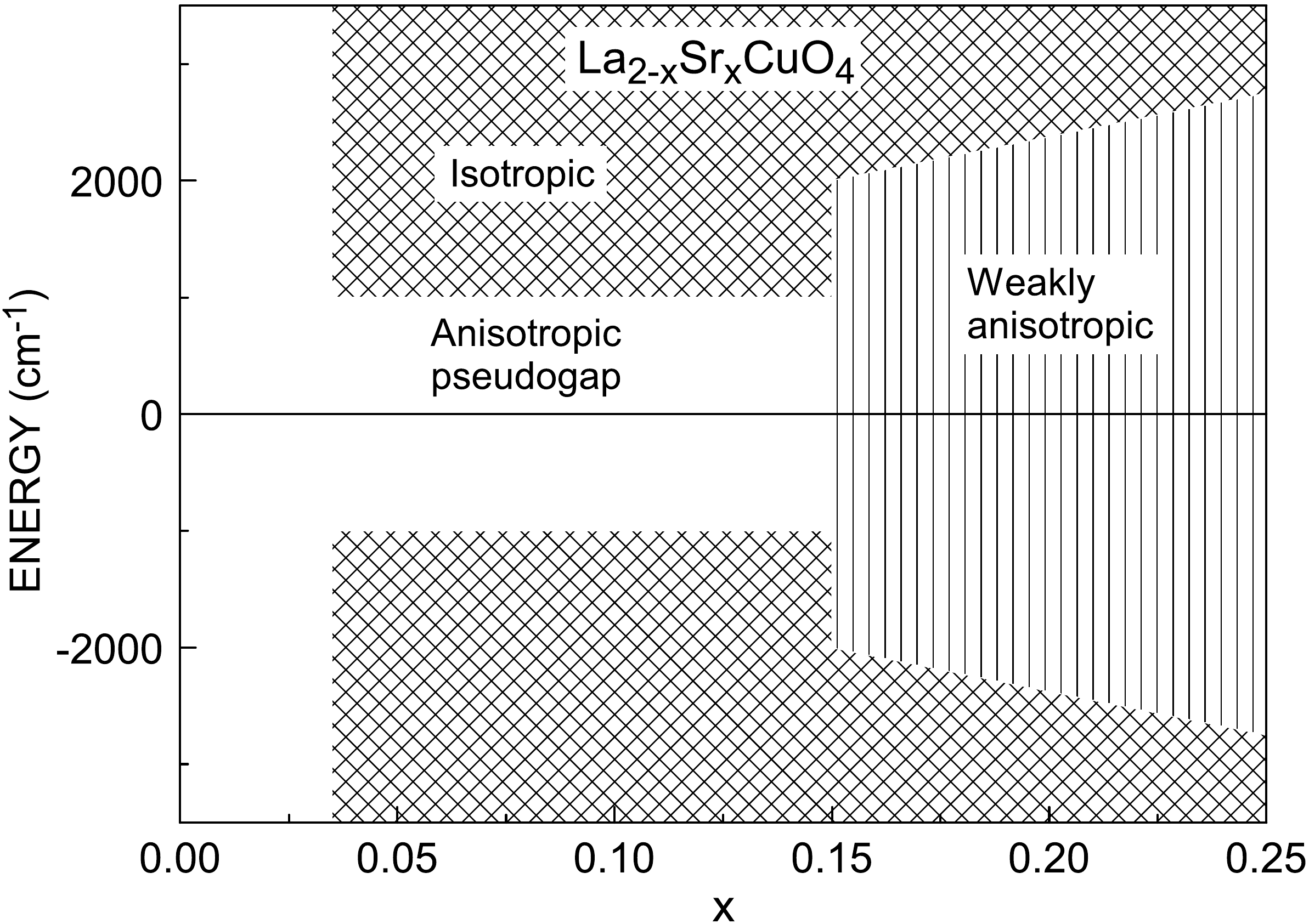}
\caption{
Phase diagram of electronic states with the isotropic or anisotropic $k$ dependence.  
The state at the chemical potential ($\epsilon=0$) is anisotropic.  
The anisotropy decreases as the energy moves away from the chemical potential and 
the state is smoothly connected to the isotropic state.  
Note that the boundary is smooth.  
}
\end{center}
\end{figure}
The isotropic and anisotropic regions in the $k$ space obtained from the $B_{\rm 1g}$ 
and $B_{\rm 2g}$ spectra are shown in Fig. 8, on the assumption that the electronic 
properties are symmetric with respect to the chemical potential.  
It is noted that the boundaries are continuous.  
The decrease of the $B_{\rm 1g}$ intensity below 2000 cm$^{-1}$ in the underdoped phase 
is due to the opening of the pseudogap near $(\pi,0)$ in agreement with 
ARPES \cite{Norman1998,Yoshida,Shi,Yoshida2}.  
The pseudogap observed in Raman scattering does not close at 300 K ($>T^*$).  
The opening of the pseudogap above $T^*$ is also reported in ARPES \cite{Kordyuk}.  
The electronic states at far sites more than 1000 cm$^{-1}$ from the chemical potential lose 
the selection rule between $B_{\rm 1g}$ and $B_{\rm 2g}$.  
The electronic states are isotropic in $k$ space.  
It is the same as the dynamical mean field theory that the $k$ dependence is ignored.  
In the overdoped phase the pseudogap closes and the intensity ratio of the $B_{\rm 1g}$ 
to the $B_{\rm 2g}$ spectra becomes increasingly large, as the carrier 
density increases.
The electronic states are approaching the band model.  
The similar phase diagram can be obtained from the $A_{\rm 1g}$ scattering.  
The isotropy increases as the energy goes away from the chemical potential similarly 
to Fig. 8.  
The $A_{\rm 1g}$ spectra have almost the same structure above $x=0.1$ as shown in Fig. 5, 
so that the boundary at $x=0.15$ is missing.  
The isotropic momentum dependence is also observed in YBa$_2$Cu$_3$O$_{6.5}$ above 
100 meV in neutron scattering \cite{Stock}.

\begin{figure}
\begin{center}
\includegraphics[trim=0mm 0mm 0mm 0mm, width=7cm]{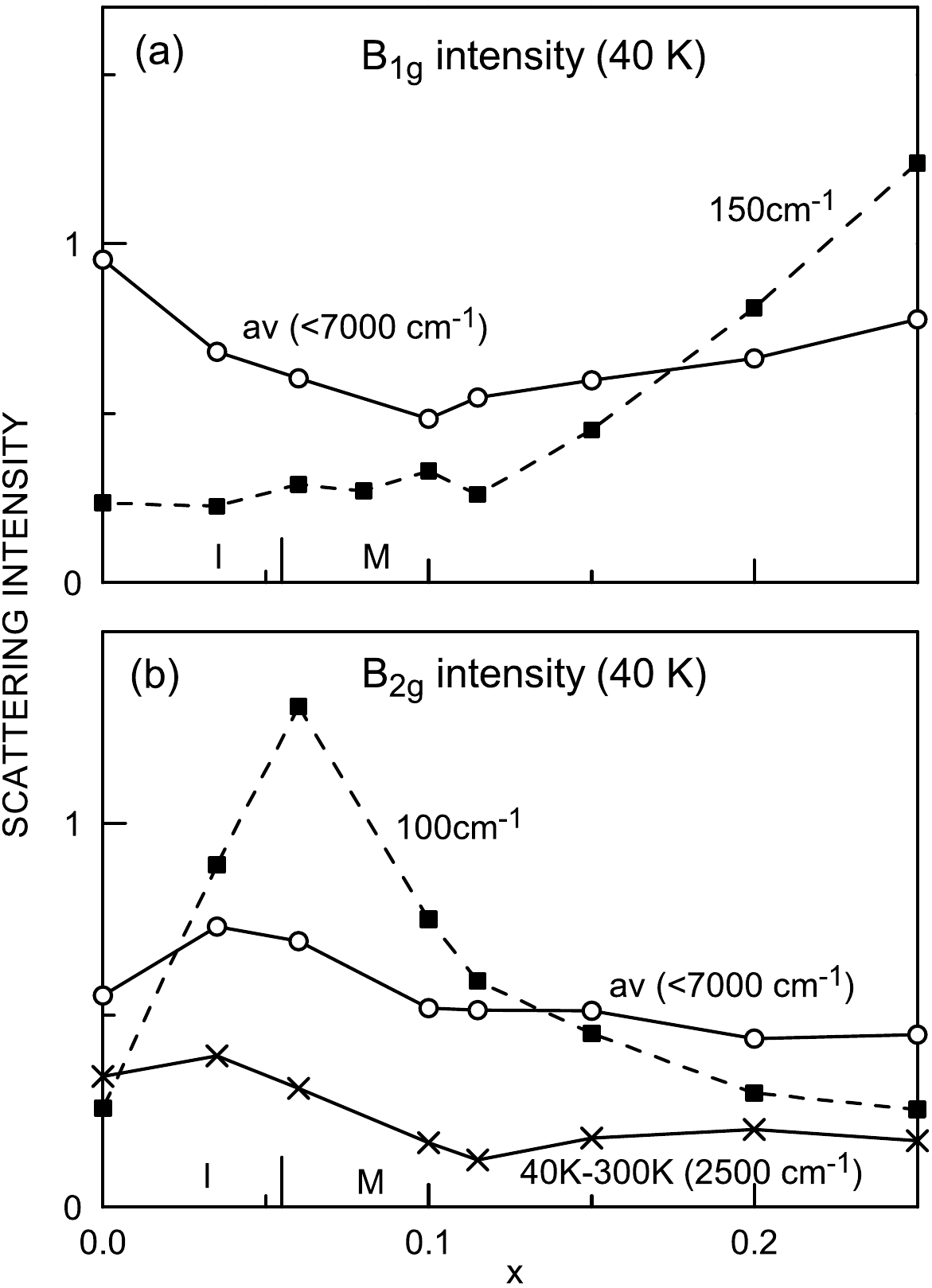}
\caption{
(a) The $B_{\rm 1g}$ low-energy intensity at 150 cm$^{-1}$ and the average intensity 
from 16 cm$^{-1}$ to 7000 cm$^{-1}$.  
(b) The $B_{\rm 2g}$ intensity at 100 cm$^{-1}$ and the average intensity 
from 16 cm$^{-1}$ to 7000 cm$^{-1}$.  
The differential spectra between 40 K and 300 K at 2500 cm$^{-1}$ are also shown.  
}
\end{center}
\end{figure}
Figure 9 shows the carrier density dependent (a) $B_{\rm 1g}$ and (b) $B_{\rm 2g}$ 
average scattering intensity from 16 to 7000 cm$^{-1}$ (solid lines).  
The $B_{\rm 1g}$ intensity decreases from $x=0$ to 0.1, because the two-magnon scattering 
intensity decreases.
The electronic scattering intensity increases as the carrier density increases.  
The $B_{\rm 2g}$ scattering intensity increases from $x=0$ to 0.06 and then gradually 
decreases with increasing the carrier density.  
The rather large average intensity at $x=0$ is due to the natural hole doping of 
our sample.  
An example of small $B_{\rm 2g}$ intensity at $x=0$ was reported \cite{Singh}.  
The $B_{\rm 2g}$ average intensity has a dip at $x=1/8$ in Fig. 9(b).  
The $B_{\rm 2g}$ spectra has a hump from 1000 to 3500 cm$^{-1}$ whose energy changes 
with the carrier density in Fig. 4(c).  
The hump is strongly enhanced as temperature decreases.  
The differential intensity at 2500 cm$^{-1}$ between 40 K and 300 K is shown in 
Fig. 9(b).  
The dip at $x=1/8$ comes from the reduction of the enhancement at low temperatures.  
The dashed lines in Fig. 9(a) and (b) show the intensity at 150 cm$^{-1}$ in the 
$B_{\rm 1g}$ spectra and 100 cm$^{-1}$ in the $B_{\rm 2g}$ spectra, respectively.  
The average intensity of the wide-energy spectra has similar carrier density dependence 
to the low-energy intensity, if two-magnon scattering is removed.  
Therefore the wide-energy electronic scattering is generated by the same mechanism 
as the low-energy scattering.  
The carrier density dependences of the low-energy $B_{\rm 1g}$ and $B_{\rm 2g}$ 
intensities are consistent with the ARPES intensities near $(\pi,0)$ and $(\pi/2, \pi/2)$, 
respectively \cite{Yoshida}.  
The fine structure is, however, different as discussed in Section~\ref{subsec:low}.  

\begin{figure}
\begin{center}
\includegraphics[trim=0mm 0mm 0mm 0mm, width=7cm]{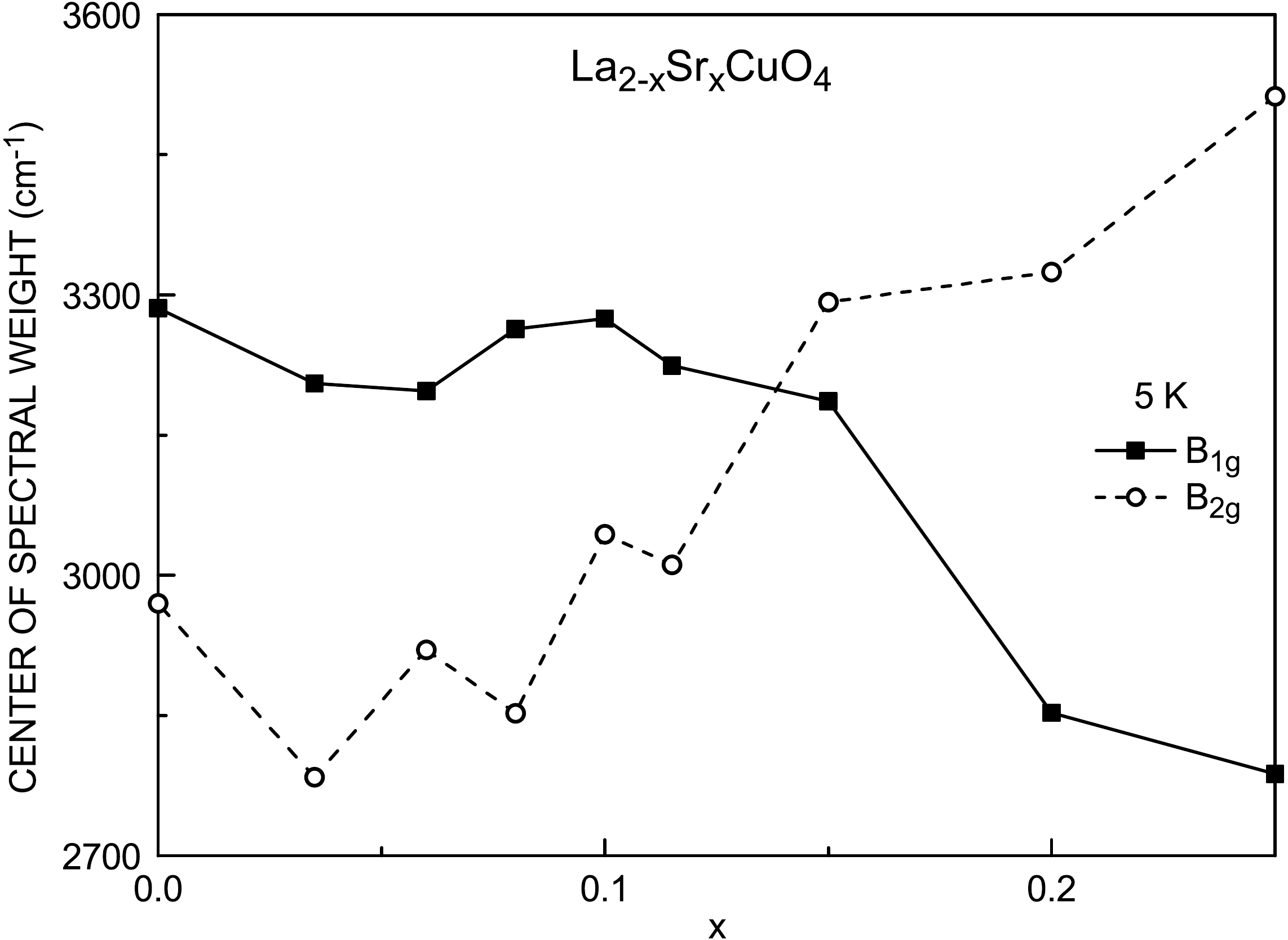}
\caption{
Center of the spectral weight below 7000 cm$^{-1}$.  
}
\end{center}
\end{figure}
Figure 10 shows the central energy of the $B_{\rm 1g}$ (solid line) and $B_{\rm 2g}$ 
(dashed line) spectral weight.  
The $B_{\rm 1g}$ central energy decreases as the carrier density increases above 
$x=0.15$.
On the other hand the $B_{\rm 2g}$ central energy increases with increasing carrier 
density.

\subsection{\label{subsec:highperp}Wide-energy spectra : $k\parallel$ and $k\!\perp$ 
stripe excitations}
We analyze the $B_{\rm 1g}$ and $B_{\rm 2g}$ spectra, because the high-energy part 
of the $A_{\rm 1g}$ spectra is strongly screened.  
The smooth $B_{\rm 2g}$ spectra at 300 K in Fig. 4(c) may be interpreted by the 
electronic Raman scattering theory with strong correlation 
\cite{Freericks2001,Freericks,Devereaux,Caprara,Kupcic}.  
However, the hump which develops from 1000 to 3500 cm$^{-1}$ as temperature 
decreases cannot be interpreted by the above models.  
The hump is isotropic and the energy depends on the carrier density.  
The enhancement of the hump on cooling is largest at 
$x = 0.035$ and smallest at $x=1/8$ in Fig. 4(c) and 9(b).  
The ``hour-glass" like magnetic susceptibility observed in neutron scattering is mainly 
analyzed by the dynamical stripes with mixed 
directions \cite{Batista,Vojta2004,Uhrig,Seibold,Vojta2006,Seibold2} 
or the interacting fermion liquid \cite{Morr,Eremin,Norman,Eremin2007}.  
We analyze the Raman spectra by individual magnetic excitations for the $k\!\perp$ 
and $k||$ stripe directions calculated by Seibold and Lorenzana \cite{Seibold,Seibold2}.

\begin{figure}
\begin{center}
\includegraphics[trim=0mm 0mm 0mm 0mm, width=7cm]{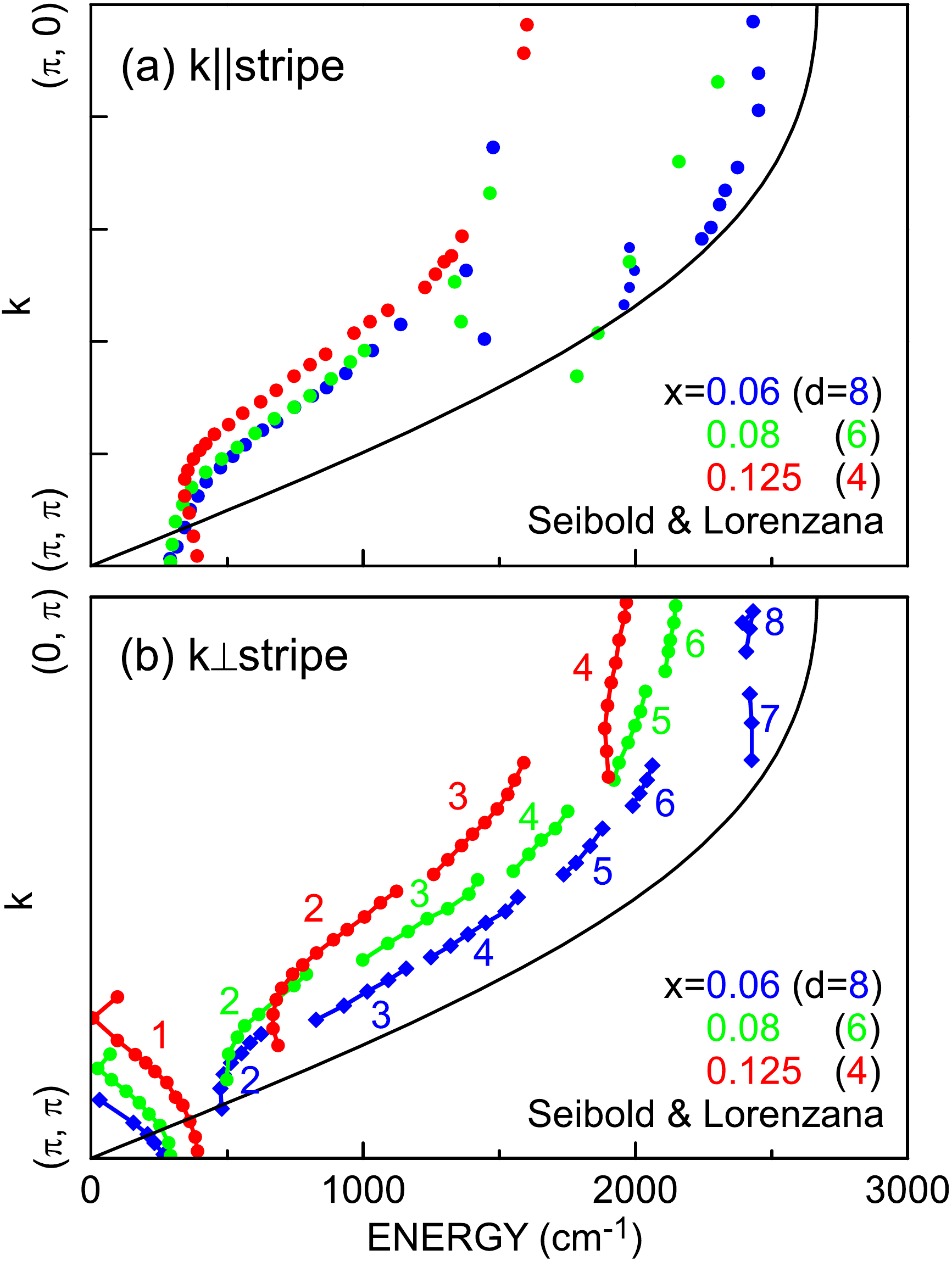}
\caption{(color online) 
$\omega \chi''(\omega, q)$ for (a) $k\parallel$ and (b) $k\!\perp$ stripe magnetic 
excitations calculated by Seibold and Lorenzana \cite{Seibold}.  
The black curve is the dispersion in the uniform antiferromagnet at $x=0$ \cite{Coldea}.
}
\end{center}
\end{figure}
Figure 11 (a) and (b) show the $\omega \chi''(\omega, q)$ for 
$k\parallel$ and $k\!\perp$ stripe in the metallic vertical 
bond-centered stripe (VBC) phase calculated by Seibold and Lorenzana 
\cite{Seibold}, respectively.  
Here $\chi''(\omega, q)$ is the imaginary part of the transverse magnetic 
susceptibility.  
The intensity representation is simplified from the original 
contour map \cite{Seibold}.  
The blue, green, and red curves represent the dispersions at 
$x = 0.06$ ($d = 8$), 0.08 ($d = 6$), and $x = 0.125$ ($d = 4$), 
respectively, where $d=\pi/\delta$ is the stripe width (inter-charge stripe 
distance) in the unit of Cu-Cu distance.  
In the $k\parallel$ stripe of Fig. 11(a) the dispersion energy rapidly 
decreases as well as the decrease of the high-energy intensity 
with increasing the carrier density.  
On the other hand in the $k\!\perp$ stripe of Fig. 11(b) the dispersion curve is 
separated into $d$ segments because of the Brillouin zone folding.  
The highest energy at $(0, \pi)$ little decreases with increasing 
the carrier density.  
The energy of each dispersion segment increases with increasing 
the carrier density from $x = 0.06$ to 0.125, because the number 
of segments decreases.  
The separated dispersion has a large energy gap between the 
first and second dispersion segments.  
At $x = 0.125$ another large gap opens between the third and 
fourth dispersion segments.  
The black line shows the uniform spin wave dispersion along the $a$ 
or $b$ axis at $x=0$ with the nearest and the next nearest neighbor exchange 
interaction energies $J = 840$ cm$^{-1}$ and 
$J'= -145$ cm$^{-1}$ \ \cite{Coldea}.  

The $B_{\rm 1g}$ two-magnon peak energy in Fig. 4(b) 
decreases with increasing the carrier density in the same way as the 
$k\parallel$ stripe magnetic excitations in Fig. 11(a).  

The $B_{\rm 2g}$ hump in Fig. 4(c) indicated by the downward triangles shifts 
from 900 - 3500 cm$^{-1}$ at $x = 0.06$ 
to 1600 - 3500 cm$^{-1}$ at $x = 0.25$.  
The triangles are numbered so that the 
energies are about twice the energy of dispersion segments in Fig. 11(b).  
The hump has the following properties.  
(1) The energy of the triangle 2 increases with 
increasing the carrier density from $x = 0.06$ to 0.115 and then 
becomes constant above $x=0.115$.  
(2) The hump develops as temperature decreases from 300 K to 5 K.  
(3) The hump is small near $x = 1/8$.  
(4) The hump is large near the insulator-metal transition.  
(5) The same hump is observed in the $B_{\rm 1g}$ spectra.  
The hump structure is observed in the $B_{\rm 1g}$ spectra of the report by Machtoub 
{\it et al}. \cite{Machtoub} at 2200 and 3100 cm$^{-1}$ at low temperatures.  

\begin{figure}
\begin{center}
\includegraphics[trim=0mm 0mm 0mm 0mm, width=8cm]{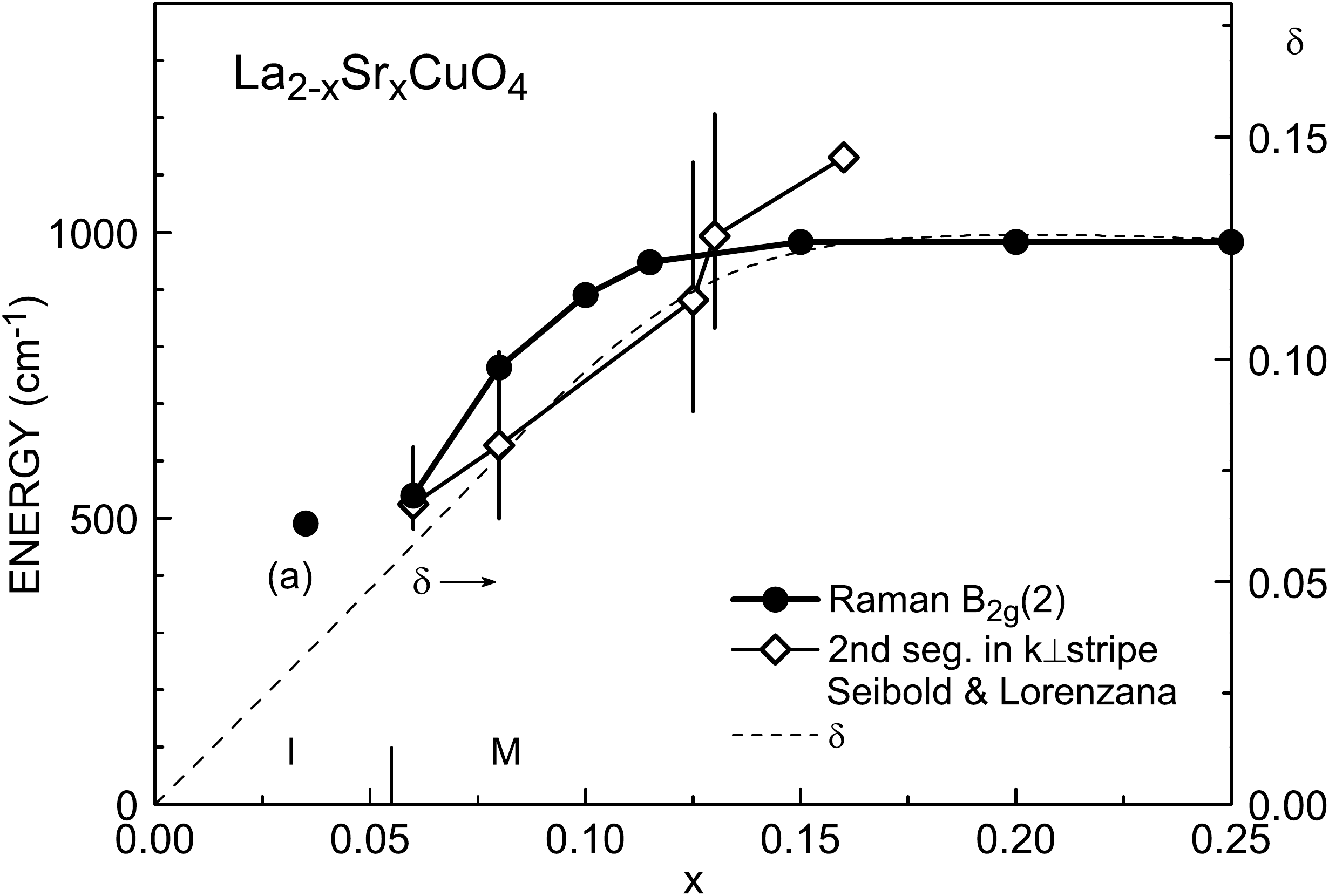}
\caption{
Edge energy of the $B_{\rm 2g}$ hump and the energy of the second 
dispersion segment in the $k\!\perp$ stripe magnetic excitations in Fig. 11 \cite{Seibold}.  
The vertical bar is the full width of the segment.  
The incommensurability $\delta$ obtained from neutron scattering is shown \cite{Yamada}.  
}
\end{center}
\end{figure}
Figure 12 shows the comparison between a half the energy of the edge 2 in the
 $B_{\rm 2g}$ spectra and the energy of the second dispersion segment in the $k\!\perp$ 
stripe excitations in Fig. 11(b) calculated by Seibold and Lorenzana \cite{Seibold}.  
The vertical bar is the energy width of the segment.  
In the metallic phase the energy 2 increases in accordance with the calculated energy 
of the second dispersion segment from $x=0.035$ to $x =0.115$.  
Above $x = 0.115$ the energy 2 remains constant, while the calculated energy keeps 
increasing.  
The incommensurability $\delta$ obtained from neutron scattering \cite{Yamada} 
is shown by the dashed line in Fig. 12.  
The $\delta$ has the similar carrier density dependence to the energy 2 of the 
present experiment.  
The saturation above $x = 1/8$ might be related to the recent Compton scattering 
that the excess hole orbital populates in Cu $d$ $3z^2-r^2$ besides O $p$ in the 
overdoped phase \cite{Sakurai}.  

Thus we conclude that the $B_{\rm 1g}$ spectra have the $k||$ and $k\!\perp$ 
stripe excitations and the $B_{\rm 2g}$ spectra have $k\!\perp$ stripe excitations.  
The electronic scattering has only $k\!\perp$ stripe component.  
The results are summarized in Table~\ref{tab:table1}.

\subsection{\label{subsec:low}Low-energy spectra : Polaron and SDW/CDW gap}
\begin{figure*}
\begin{center}
\includegraphics[trim=0mm 0mm 0mm 0mm, width=15cm]{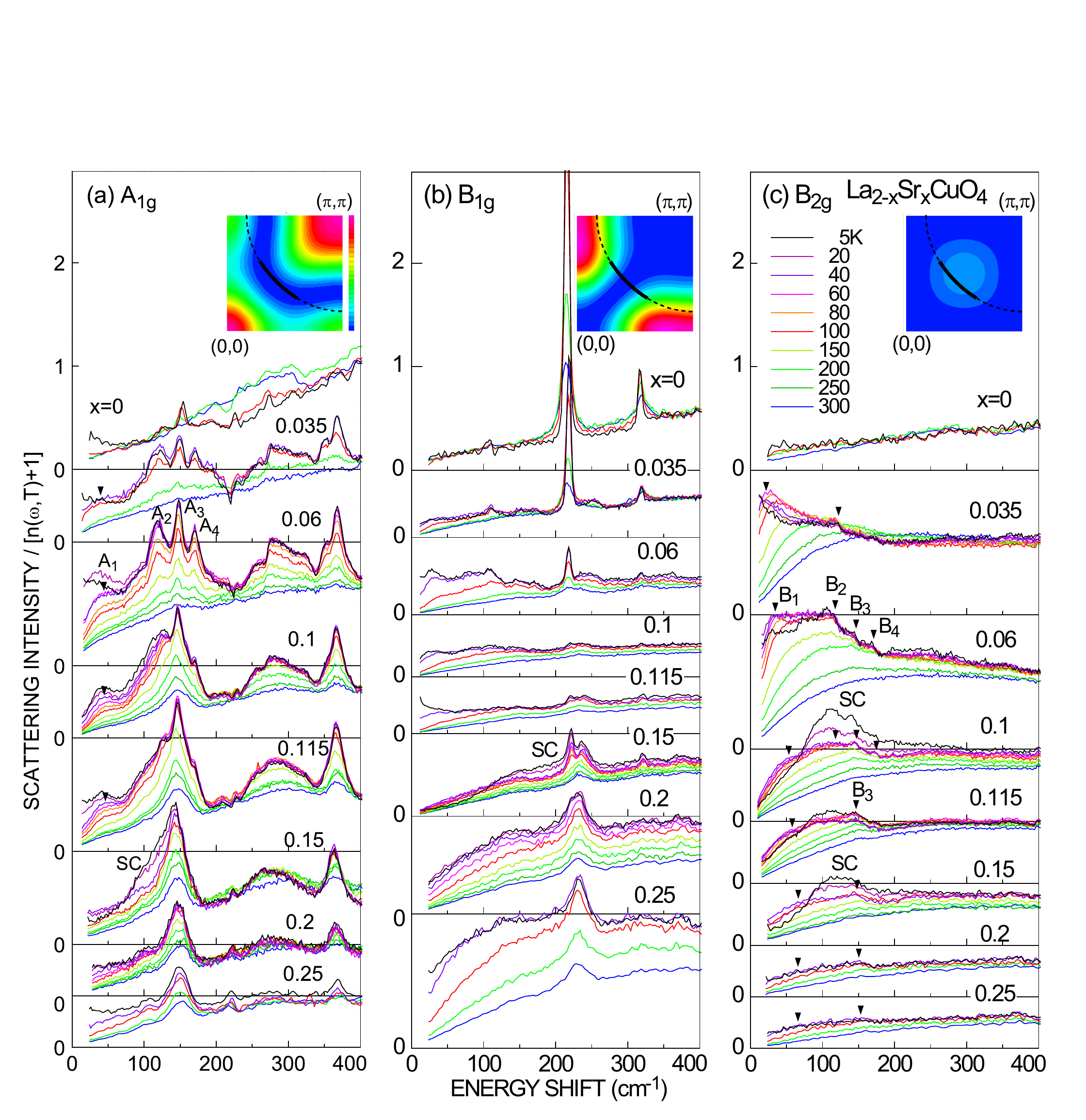}
\caption{(color online) 
(a) $A_{\rm 1g}$, (b) $B_{\rm 1g}$ and (c) $B_{\rm 2g}$ low-energy Raman spectra.  
The insets show the contour maps of $(1/m^*)^2$ for $A_{\rm 1g}$, $B_{\rm 1g}$ 
and $B_{\rm 2g}$.  
}
\end{center}
\end{figure*}
The low-energy spectra are different depending on the symmetry.  
The $B_{\rm 1g}$ spectra observe the antinodal gap near $(\pi,0)$ and the $B_{\rm 2g}$ 
spectra observe the nodal gap near $(\pi/2, \pi/2)$ in accordance with the tight binding 
band model of Eq.~(\ref{eq:tightbinding}) \cite{Devereaux1994,Devereaux1995}.  
The $B_{\rm 1g}$ and $B_{\rm 2g}$ superconducting gaps were detected experimentally 
\cite{Opel,Sugai,Tacon,Muschler,Munnikes}.  
The absorption coefficient corrected low-energy spectra are shown in Fig. 13.  
The insets show the contour maps of 
$|\partial ^2\epsilon({\bf k})/\partial k_{\alpha} \partial_{\beta}|^2$ in Fig. 1.  
The absorption uncorrected spectra were presented in the previous paper \cite{Sugai3}.  
The $B_{\rm 1g}$ and $B_{\rm 2g}$ spectra are similar to other groups \cite{Muschler}, 
but our spectra have finer structure because all the spectra were obtained on 
fresh cleaved surfaces.  

The structural transition temperature from the tetragonal $I4/mmm$ to orthorhombic $Cmca$ 
phase decreases from 525 K at $x=0$ to 10 K at $x=0.21$ \cite{Keimer1992,Radaelli}.  
The orthorhombic crystallographic axes $a$ and $b$ rotate by $45^{\circ}$ from the 
tetragonal axes $x$ and $y$ and the unit cell volume doubles.  
The optical phonon modes are $2A_{\rm 1g}+2E_{\rm g}+3A_{\rm 2u}+B_{\rm 2u}+4E_{\rm u}$ 
in the tetragonal structure and 
$5A_{\rm g}+3B_{\rm 1g}+6B_{\rm 2g}+4B3_{\rm g}+4A_{\rm u}+6B_{\rm 1u}+4B_{\rm 2u}+7B_{\rm 3u}$ 
in the orthorhombic structure.  
The $(\pi,\pi)$ points in the tetragonal structure becomes the $\Gamma$ point in the 
orthorhombic structure.  
The selection rule viewed from the tetragonal axes is listed in Table~\ref{tab:table2}.  
\begin{table}
\caption{\label{tab:table2}Selection rule for phonon modes.  
}
\begin{ruledtabular}
\begin{tabular}{c | c c c}
Symmetry in $D_{4h}$&$A_{\rm 1g}$&$B_{\rm 1g}$&$B_{\rm 2g}$\\
\hline
Polarization&$[(xx)+(aa)$&$(xy)$&$(ab)$\\
in $D_{4h}$&$-(xy)-(ab)]/2$& & \\
\hline
Tetragonal$(D_{4h})$&$2A_{\rm 1g}$&0&0\\
Orthorhombic$(D_{2h})$&$5A_{\rm g}$&$3B_{\rm 1g}$&$5A_{\rm g}$\\
\end{tabular}
\end{ruledtabular}
\end{table}

The $A_{\rm 1g}$ and $B_{\rm 2g}$ spectra are rapidly enhanced as carriers are doped, while 
the $B_{\rm 1g}$ spectra are weak.  
The $A_{\rm 1g}$ and $B_{\rm 2g}$ low-energy spectra are strongly enhanced as temperature 
decreases in the underdoped phase.  
The intensities decrease in the overdoped phase and the $B_{\rm 1g}$ spectra 
becomes strong instead.  

\begin{figure}
\begin{center}
\includegraphics[trim=0mm 0mm 0mm 0mm, width=8cm]{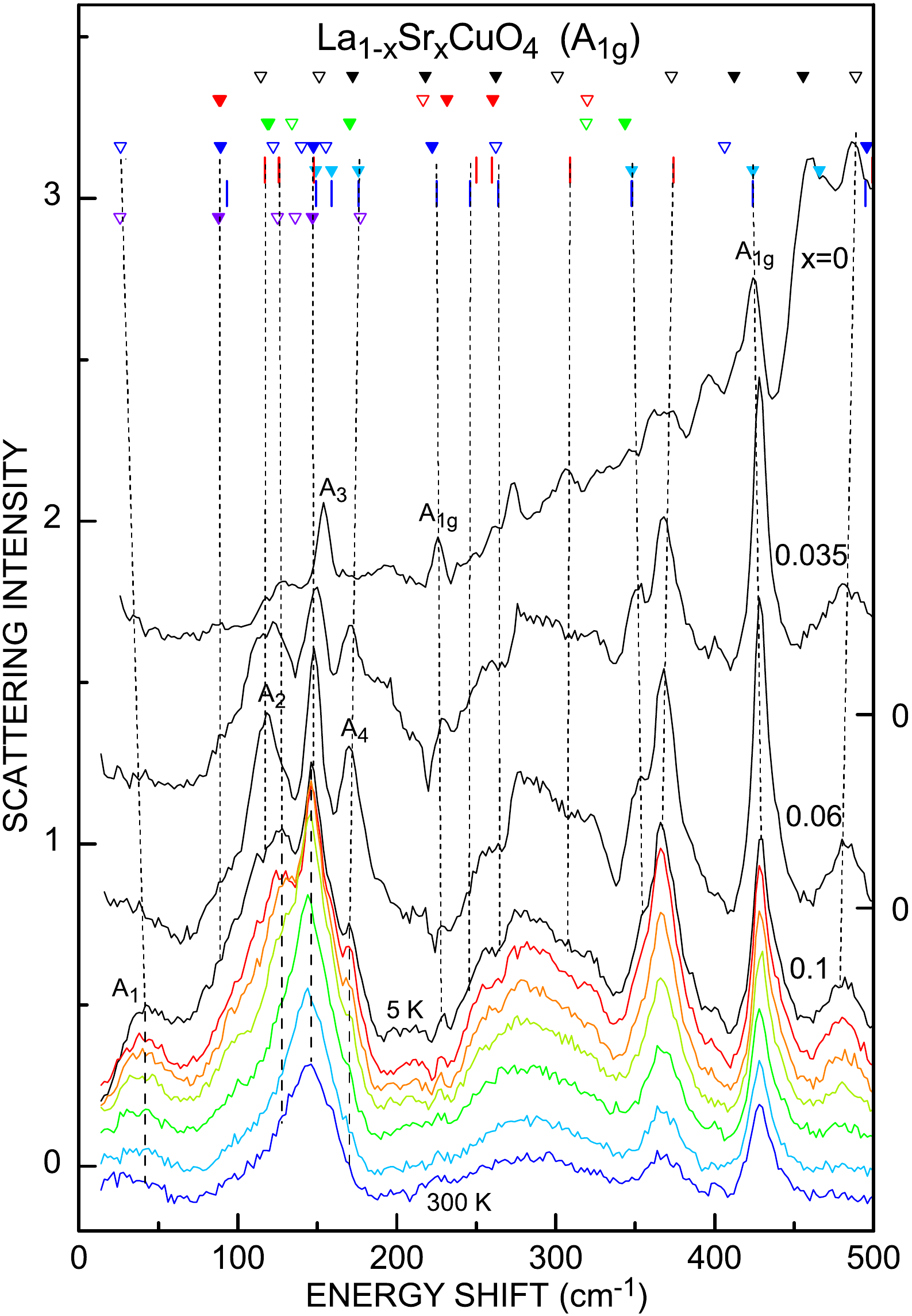}
\caption{(color online) 
The $A_{\rm 1g}$ Raman spectra at $x=0$, 0.035, and 0.6 at 5 K and 0.1 from 5 (black), 
40 (red), 100 (orange), 150 (yellowish green), 200 (green), 250 (light blue) to 300 K (blue).  
The black, red, green, and blue triangles in the upper four lines are $\Sigma_1-\Sigma_4$ 
modes at ${\bf q}=(0,0)$ (filled) and $(\pi,\pi)$ (open) of the tetragonal structure, 
respectively \cite{Chaplot}.  
The light blue triangles and red short bars on the fifth line are $\Sigma_1$ modes 
at $(0,0)$ and $(\pi,\pi)$, respectively \cite{Rietschel1989}.  
The blue short bars on the sixth line are $\Lambda$ modes at $(0,0)$ \cite{Rietschel1989}.  
The purple triangles in the seventh line are $\Sigma$ modes at $(0,0)$ (filled) 
and $(\pi,\pi)$ (open), respectively \cite{Birgeneau1987,Boni}.  
The zero levels for the spectra at $x=0.06$ and 0.1 (5 K) are shown in the left scale.  
The higher-temperature spectra are downward shifted by 0.1 in the order of increasing 
temperature.  
The zero levels for the spectra at $x=0$ and 0.035 are shown in the right scale.  
}
\end{center}
\end{figure}
The $A_{\rm 1g}$ spectra in Fig. 13 have many phonon peaks.  
Many of them are derived from Raman inactive modes.  
Two $A_{\rm 1g}$ phonon modes in the tetragonal structure have the atomic displacements 
in the $c$ direction.  
Therefore the Raman intensity is strong in the (c,c) polarization.  
The energies are 229 and 426 cm$^{-1}$ at 5 K and $x=0$ \cite{Sugai1989}.  
The $A_{\rm 1g}$ intensities in the in-plane polarization spectra are weak in Fig.13(a).  
The other peaks in the (c,c) spectra are 126, 156, and 273 cm$^{-1}$ at 5 K \cite{Sugai1989}.  
The peaks activated in the orthorhombic distortion disappear at $x=0.24$ \cite{Lampakis2006}, 
because the orthorhombic structure ends at x=0.21 and 10 K \cite{Radaelli}.  

In order to find out the origin of the phonon peaks in the $A_{\rm 1g}$ spectra, 
neutron scattering results are plotted together with the Raman spectra in Fig. 14.  
The $A_{\rm 1g}$ spectra of $x=0$, 0.035, and  0.06 at 5 K and 0.1 from 5 
to 300 K are shown.  
The upper black, red, green, and blue triangles are $\Sigma_1$, $\Sigma_2$, $\Sigma_3$, 
and $\Sigma_4$ modes at ${\bf q}=(0,0)$ (filled) and $(\pi,\pi)$ (open) of the 
tetragonal structure, respectively \cite{Chaplot}.  
The light blue triangles and red short bars on the fifth line are $\Sigma_1$ modes 
at $(0,0)$ and $(\pi,\pi)$, respectively \cite{Rietschel1989}.  
The blue short bars on the sixth line are the $\Lambda$ modes at $(0,0)$ 
\cite{Rietschel1989}.  
The purple triangles are $\Sigma$ modes at $(0,0)$ (filled) and $(\pi,\pi)$ (open) 
\cite{Birgeneau1987,Boni}.  
The peaks denoted by $A_{\rm 1g}$ are derived from the tetragonal $A_{\rm 1g}$ phonons.  
Many peaks can be assigned to the phonon modes observed in neutron scattering.  
The A$_1$ peak is assigned to the $\Sigma_4$ soft mode at $(\pi,\pi)$ which induces the 
tetragonal-orthorhombic structural phase transition \cite{Birgeneau1987,Boni}.  
The A$_1$ peak becomes very small at $x=0.15$.  
Only a small hump is observed at 60 K.  
The A$_1$ peak disappears at $x\ge 0.2$.  
The 88 cm$^{-1}$ hump is assigned to the same branch at (0,0).  
The A$_2$ peak is assigned to the zone boundary $(\pi,\pi)$ modes of the longitudinal 
acoustic mode ($\Sigma_1$, 125 cm$^{-1}$) and the transverse acoustic mode ($\Sigma_3$, 
136 cm$^{-1}$).  
The A$_3$ peak is assigned to the $\Sigma_1$ mode of 147 cm$^{-1}$ at (0,0) or 148 cm$^{-1}$ 
at $(\pi,\pi)$.  
The A$_4$ peak is assigned to the $\Sigma_1$  branch at $(\pi,\pi)$ (177 cm$^{-1}$).  
The A$_2$ and A$_4$ peak intensities decrease faster than the A$_3$ peak, as temperature 
increases.  

The A$_2$, A$_3$, and A$_4$ peaks are observed in infrared spectroscopy as $B_{\rm 3u}$ 
modes of the orthorhombic structure \cite{Padilla}.  
The orthorhombic crystal structure $Cmca$ has inversion symmetry, so that the Raman 
and infrared activities are exclusive.  
The appearance in both spectra means the disappearance of the inversion symmetry.  
The modes are not the simple phonons, but may be local modes which have lower symmetry 
than the macroscopic orthorhombic symmetry.  
In fact the $A_{\rm 1g}$ spectra are strongly enhanced as carriers are doped and as 
temperature decreases.  
Those $A_{\rm 1g}$ modes are not the pure phonon modes, but electron-phonon coupled modes.  

The correspondence between the $240-330$ cm$^{-1}$ peak energies and the phonon energies 
obtained from neutron scattering is not good as shown in Fig. 13 and 14, so that the peaks 
are assigned to the second order of the peaks $A_2- A_4$ and the humps near those peaks.  

Zhou {\it et al}. \cite{Zhou2005} observed multiple phonon spectra of about 17 meV 
(140 cm$^{-1}$) on the electron dispersion along the $(0,0)-(\pi,\pi)$ nodal direction 
in ARPES of underdoped LSCO.  
The energy is just the same as the average energy of the peaks A$_2$, A$_3$, and A$_4$.  
The energy resolution in ARPES is 12 and 20 meV, while that of Raman scattering is 
0.7 meV.  
Therefore the Raman scattering presents the fine structure of the electron-phonon 
coupled modes.  
The difference from the ARPES is that the first order peaks are stronger than the 
second order peaks in Raman scattering, while the higher order peaks are stronger 
than the first order peaks in ARPES \cite{Zhou2005}.  
The multiple phonon spectra are produced by the electronic scattering through 
the self-energy of multiple phonon component \cite{Shen2004,Mahan}.  
The electron-phonon coupling is more clearly observed in the $B_{\rm 2g}$ channel.  

\begin{figure}
\begin{center}
\includegraphics[trim=0mm 0mm 0mm 0mm, width=8cm]{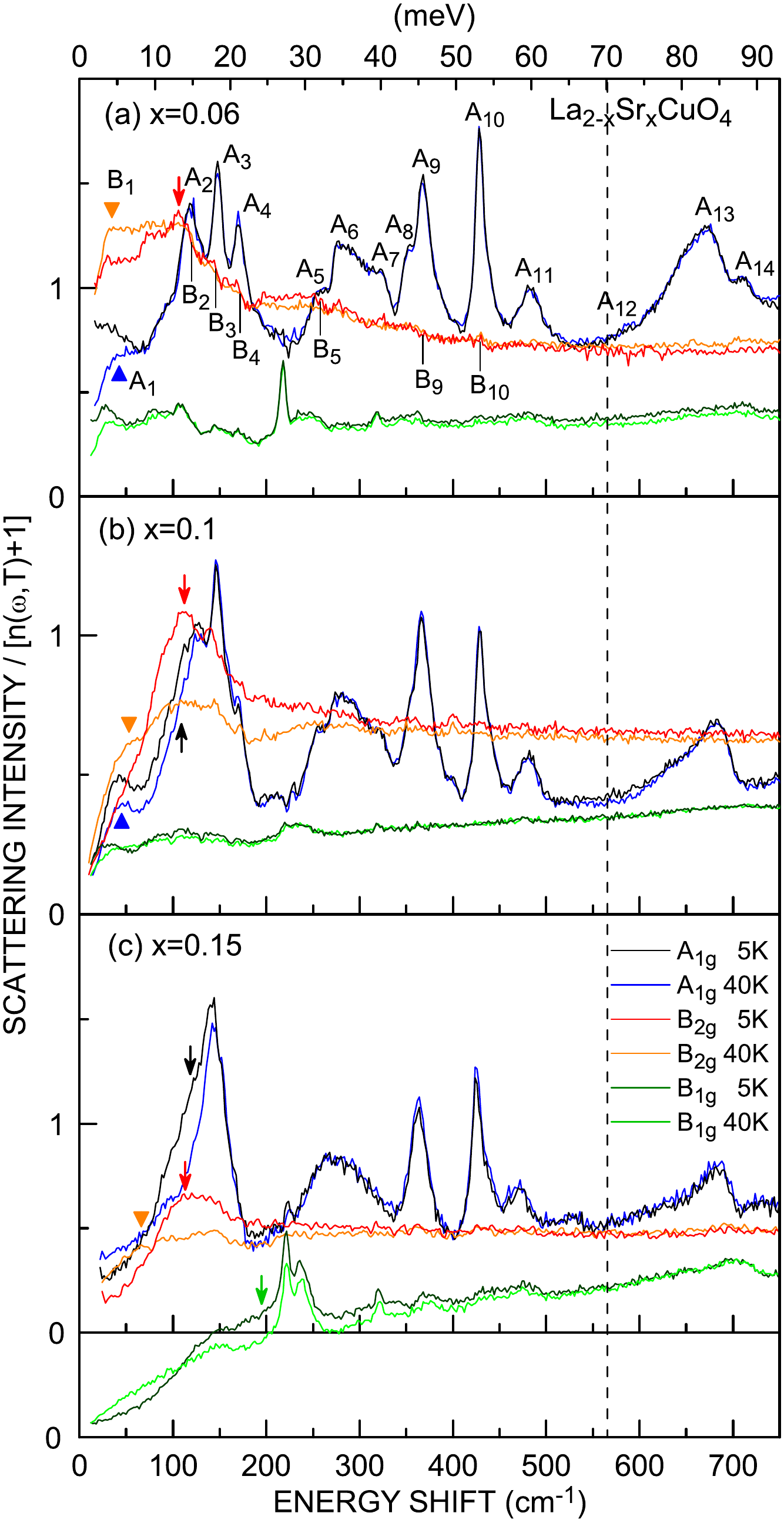}
\caption{(color online) 
$A_{\rm 1g}$, $B_{\rm 1g}$, and $B_{\rm 2g}$ low-energy spectra at 5 and 40 K.  
The arrow indicates the superconducting gap obtained from the differential spectra 
between 5 k and 40 K in Fig. 19.  
The orange triangle is the SDW/CDW gaps in the $B_{\rm 2g}$ spectra at 40 K.  
The blue triangle is the lowest-energy peak in the $A_{\rm 1g}$ spectra at 40 K.  
}
\end{center}
\end{figure}
\begin{figure}
\begin{center}
\includegraphics[trim=0mm 0mm 0mm 0mm, width=8cm]{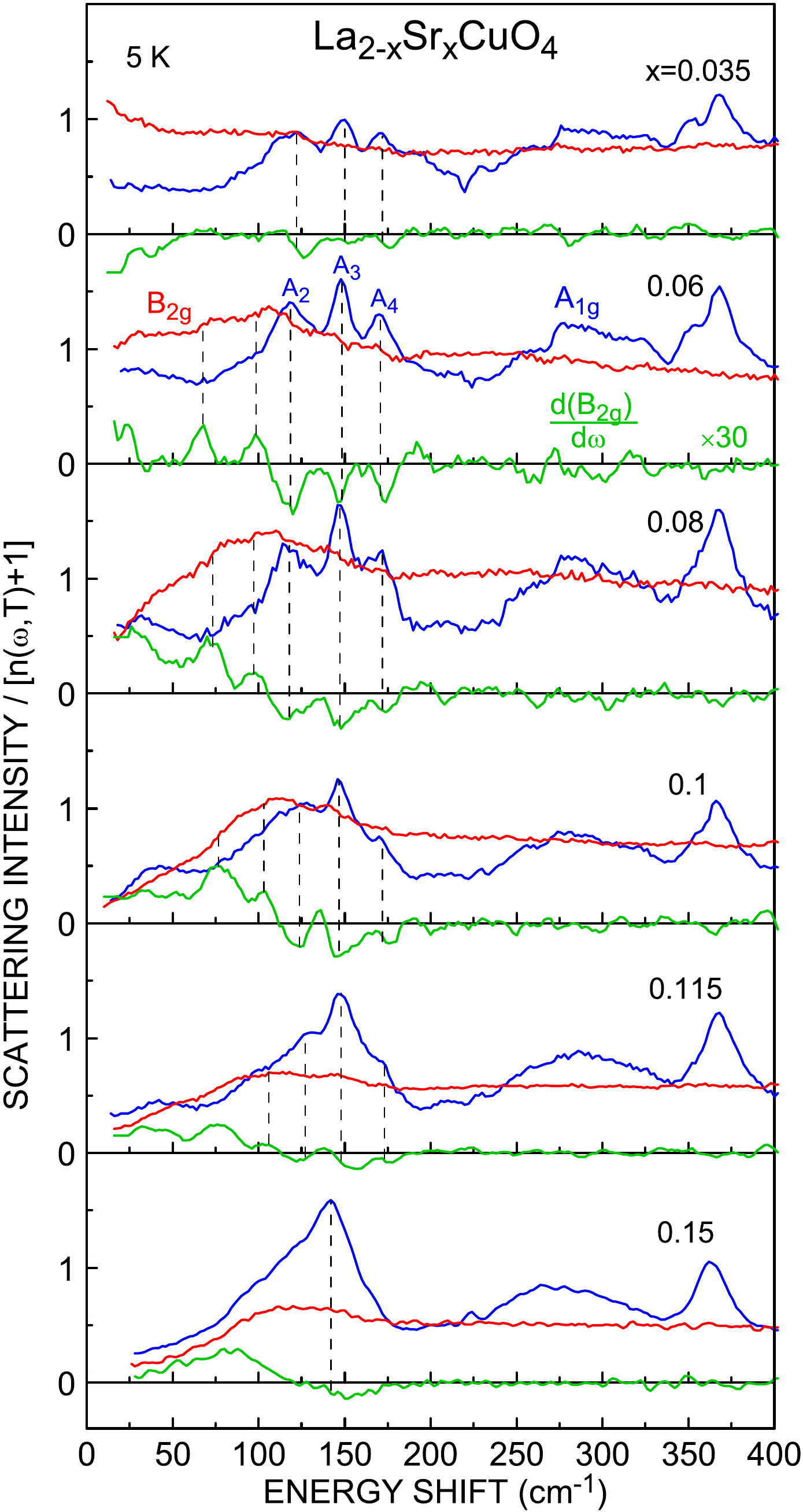}
\caption{(color online) 
Correspondence between the $A_{\rm 1g}$ (Blue) spectra and the derivative of the $B_{\rm 2g}$ 
(red) spectra, $d(B_{\rm 2g})/d\omega$ (green).  
The A$_2$, A$_3$, and A$_4$ peaks correspond to the minima of $d(B_{\rm 2g})/d\omega$.  
The $d(B_{\rm 2g})/d\omega$ is plotted with 30 times enlarged intensity.  
}
\end{center}
\end{figure}
The $B_{\rm 2g}$ spectra at 300 K in Fig. 13(c) are strongly enhanced by the small carrier 
doping of $x=0.035$ even in the insulating phase.  
The low-energy part below 180 cm$^{-1}$ is further enhanced at $x\ge 0.035$ as 
temperature decreases.  
Figure 15 shows the comparison among the $A_{\rm 1g}$ (black and blue), $B_{\rm 1g}$ 
(dark green and green) and $B_{\rm 2g}$ (red and orange) spectra at 5 K and 40 K.  
The $B_{\rm 2g}$ peak below 180 cm$^{-1}$ has the steps B$_2$, B$_3$, and B$_4$ as denoted 
in the spectra of $x=0.06$ in Fig. 15(a).  
The energies of the peaks A$_2$, A$_3$, and A$_4$ are the same as the energies of steps 
B$_2$, B$_3$, and B$_3$.  
It is more clearly observed by taking the derivative of the $B_{\rm 2g}$ spectra with 
respect to the energy.  
Figure 16 shows the  $A_{\rm 1g}$ (blue), $B_{\rm 2g}$ (red), and the $d(B_{\rm 2g})/d\omega$ 
(green).  
The A$_2$, A$_3$, and A$_4$ peaks correspond to the minima of the $d(B_{\rm 2g})/d\omega$ 
from $x=0.035$ to 0.15.  
It proves that the step structure in the $B_{\rm 2g}$ spectra are produced by the 
Fano resonance between the continuum electronic scattering and the sharp phonon peaks.  
It is the clear evidence that the states below 180 cm$^{-1}$ are electron-phonon coupled 
polaronic states.  
The $B_{\rm 2g}$ hump from 180 to 380 cm$^{-1}$ is the second order of the peak from 30 to 
180 cm$^{-1}$.  
The steps are also observed at B$_5$, B$_9$, and B$_{10}$ in Fig. 15(a) which have 
the same energies of the peaks A$_5$, A$_9$, and A$_{10}$, respectively.  

ARPES observed a kink at 70 meV on the electronic dispersion in the nodal direction 
\cite{Lanzara,Zhou2003,Zhou2007}.  
It is assigned to the coupling with the half-breathing phonon mode \cite{Ishihara}.  
The A$_{13}$ peak in Fig. 15 is derived from the $\Gamma$ point mode of the highest 
$\Delta_1$ and $\Sigma_1$ longitudinal phonon branch.  
The small hump A$_{12}$ is the half-breathing mode which is the $(\pi,0)$ mode of 
the $\Delta_1$ branch \cite{McQueeney,Pintschovius,McQueeney2001,Pintschovius2006}.  
No structure is observed in the $B_{\rm 2g}$ spectra at 70 meV.  
The A$_{14}$ peak is the breathing mode which is the $(\pi,\pi)$ mode of the 
$\Sigma_1$ branch.  

The $B_{\rm 2g}$ intensity at 100 cm$^{-1}$ is shown in Fig. 9(b) as a 
representative of the low-energy peak which is enhanced at low temperatures.  
The intensity rapidly increases from $x=0$ to 0.06 and then gradually decreases 
with increasing the carrier density.  
It is consistent with the ARPES intensity near $(\pi/2, \pi/2)$ \cite{Yoshida}.  
However, it contradicts to the calculation that the $B_{\rm 2g}$ intensity is small 
as discussed in Section~\ref{subsec:eleram} \cite{Shvaika2005}.  
The formation of polaronic states may be the origin of the large scattering intensity 
at low temperatures.
It is discussed in Section~\ref{sec:pseudogap}.  

\begin{figure}
\begin{center}
\includegraphics[trim=0mm 0mm 0mm 0mm, width=8cm]{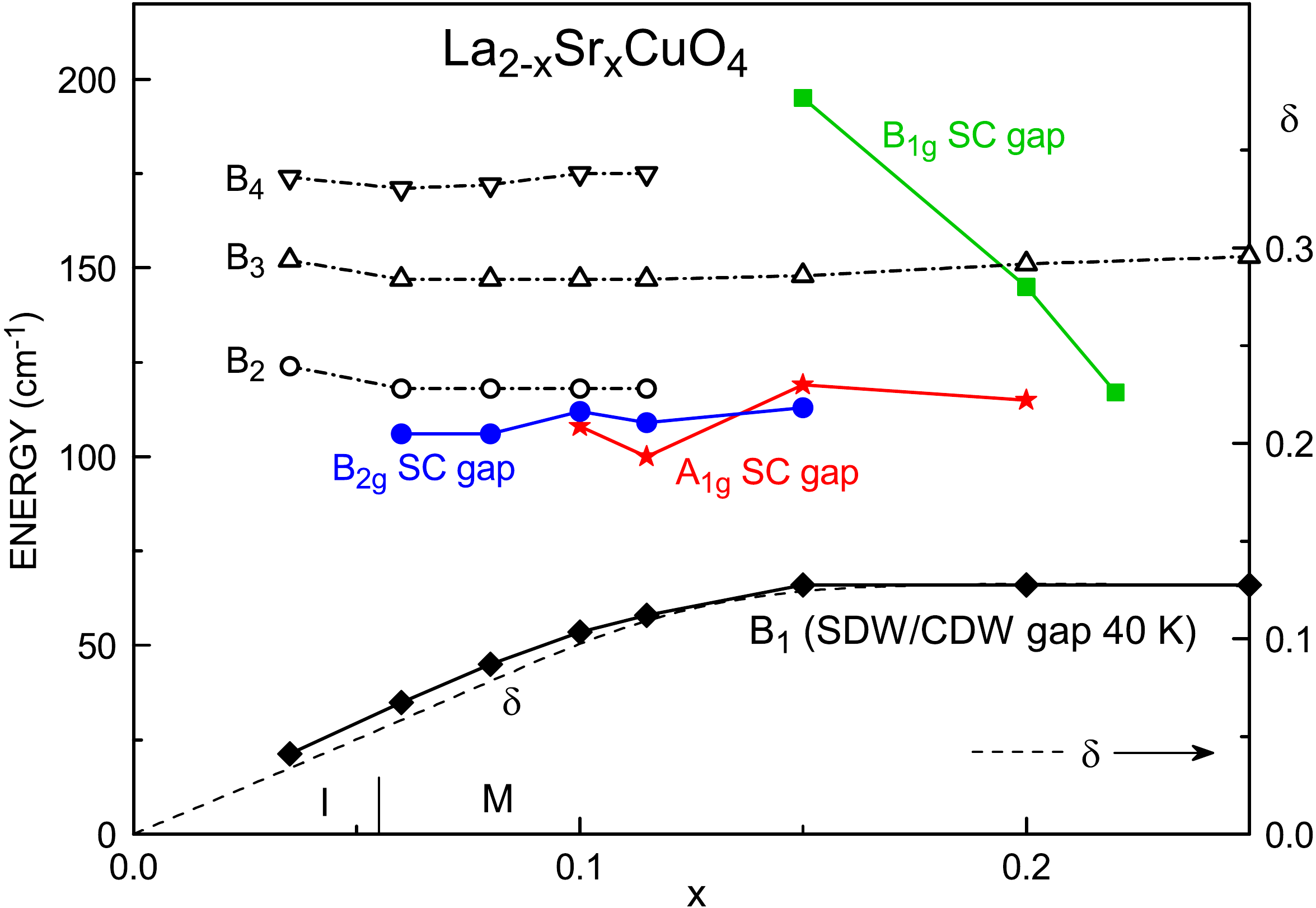}
\caption{(color online) 
Solid symbols: $A_{\rm 1g}$, $B_{\rm 1g}$, and $B_{\rm 2g}$ pair breaking peak energies 
($2\Delta$) at 5 K and $B_{\rm 2g}$ SDW/CDW gap energy $E_1$ at 40 K. 
Open symbols: electron-phonon coupled mode energies $B_2$, $B_3$, and $B_4$ at 40 K.  
Dashed curve: the incommensurability $\delta$ \cite{Yamada}.  
}
\end{center}
\end{figure}
The A$_1$ peak in Fig. 13(a) and 14 is derived from the intrinsic $A_{\rm g}$ mode in the 
orthorhombic $Cmca$ structure.  
This mode is the soft mode inducing the 
tetragonal-orthorhombic phase transition \cite{Birgeneau1987,Boni}.  
The A$_1$ peak energy at 40 K and $x=0.035$ is 39 cm$^{-1}$, while the B$_1$ peak energy in 
Fig. 13(c) is 21 cm$^{-1}$ at $x=0.035$.  
The A$_1$ peak energy does not decrease on approaching $x=0$, because the 
tetragonal-orthorhombic transition temperature increases \cite{Keimer1992,Radaelli}.  
On the other hand the B$_1$ peak energy decreases as $x$ decreases.  
Therefore the origin of the B$_1$ peak is different from the A$_1$ peak. 
The $B_{\rm 2g}$ low-energy intensity increases at $x=0.035$, as temperature decreases 
to 60 K and then the intensity below 70 cm$^{-1}$ decreases at 40 K.  
The temperature for the intensity drop below 70 cm$^{-1}$ decreases to 5 K at 
$x=0.06$ \cite{Sugai2}.  
The low energy side steeply decreases to make a gap at $x=0.035$ and 0.06 in Fig. 13(c).  
The gap is partially buried and the metallic conductivity is achieved 
at $x\ge 0.6$.  
The B$_1$ peak or edge becomes weak at $x\ge 0.2$, but the kink can be observed, 
when the intensity scale is magnified.  
Figure 17 shows the carrier density dependence of the B$_1- B_4$ peak energies and the 
incommensurability $\delta$ (dashed line) obtained from the neutron 
scattering spots $(\pi \pm \delta, \pi)$ and $(\pi, \pi \pm \delta)$ \cite{Yamada}.   
The B$_1$ energy increases as the carrier density increases from $x=0.035$ to 1/8 and 
then becomes constant in good accordance with $\delta$.  
Therefore $B_1$ is assigned to the SDW/CDW gap.  

In the $B_{\rm 1g}$ spectra of Fig. 13(b) the 216 and 317 cm$^{-1}$ peaks at $x=0$ are 
intrinsic $B_{\rm 1g}$ phonon peaks in the orthorhombic structure.  
The $B_{\rm 1g}$ electronic scattering presents the charge excitations near 
$(\pi, 0)$, if the Fermi surface is complete.  
However, the Fermi surface is depleted near $(\pi,0)$ due to the opening of the 
pseudogap in the underdoped phase.  
It decreases the low-energy scattering intensity below 2000 cm$^{-1}$ as stated with 
respect to Fig. 7.  
The low-energy intensity increases at $x\ge 0.15$ in accordance with the increase of the ARPES 
intensity near $(\pi, 0)$ \cite{Yoshida}.  
The carrier density dependent intensity of the representative point of 150 cm$^{-1}$ 
is shows by the dashed line in Fig. 9(a).

\begin{figure}
\begin{center}
\includegraphics[trim=0mm 0mm 0mm 0mm, width=7cm]{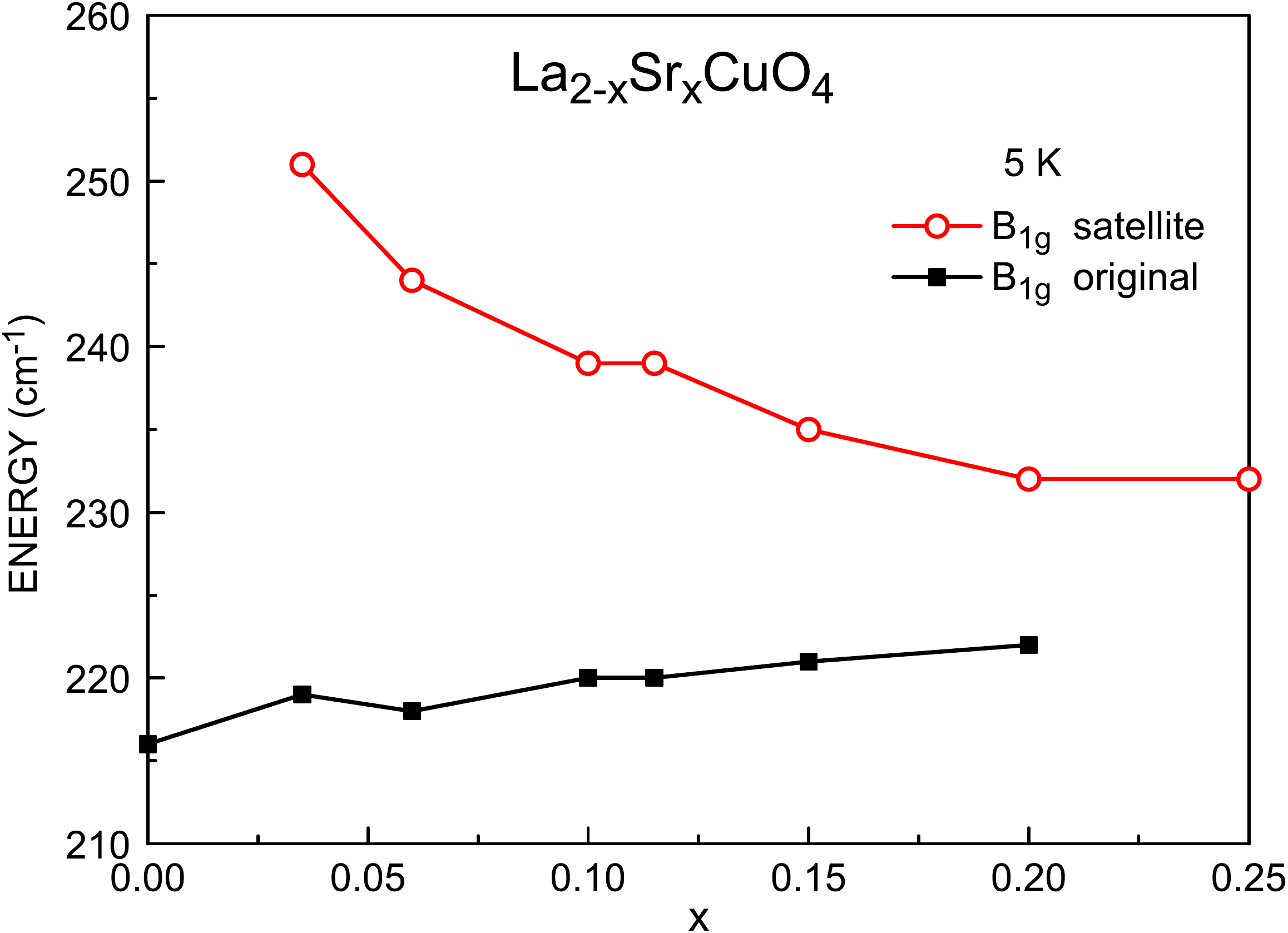}
\caption{(color online) 
Peak energies of the $B_{\rm 1g}$ original phonon mode and the satellite phonon mode.  
}
\end{center}
\end{figure}
The coupling between the $A_{\rm 1g}$ phonons and the $B_{\rm 1g}$ continuum spectra 
is weak in the overdoped phase.    
On the other hand the large coupling between the $B_{\rm 1g}$ phonon in the orthorhombic 
structure and the electronic continuum states is observed.  
The sharp $B_{\rm 1g}$ phonon peak at 216 cm$^{-1}$ $(x=0)$ splits into the original sharp 
peak and the satellite broad peak at high energy side by doping.  
The satellite peak energy decreases from 251 cm$^{-1}$ at $x=0.035$ to 232 cm$^{-1}$ 
at $x=0.25$ in Fig. 18.  
The sharp peak does not appear in the infrared spectra, but the satellite peak 
appears \cite{Padilla}.  
The intensity of the sharp peak moves into the satellite peak as carrier density increases.  
The satellite peak becomes much stronger than the original peak at $x=0.25$.  
The sum of two peak intensities decreases from $x=0$ to 0.1 and then increases at 
$x\ge 0.15$ as the electronic continuum intensity increases.  

In the crystal with inversion symmetry such as the orthorhombic $Cmca$ the Raman active 
phonon mode has even parity and the infrared active phonon mode has odd parity.  
The Raman active mode does not interact with the long wavelength plasma, so that it is 
not affected by the carrier doping.  
On the other hand the infrared active mode interacts with the plasma.  
The energy of the longitudinal optical mode changes from $\omega_{\rm LO}$ to 
$\omega_{\rm TO}$ ($<\omega_{\rm LO}$), as the plasma energy $\omega_{\rm PL}$ exceeds 
$\omega_{\rm LO}$.  
If crystal loses the inversion symmetry, some of the Raman active modes become 
infrared active.  
However, the higher energy shift of the satellite mode cannot be explained by the coupled 
mode, even if the 218 cm$^{-1}$ $(x=0.06)$ mode becomes infrared active.  
The coexistence of the original peak and the satellite peak suggests the microscopic 
inhomogeneity in the crystal.  
It is discussed in Section~\ref{subsec:dislocation}.

\begin{figure*}
\begin{center}
\includegraphics[trim=0mm 0mm 0mm 0mm, width=15cm]{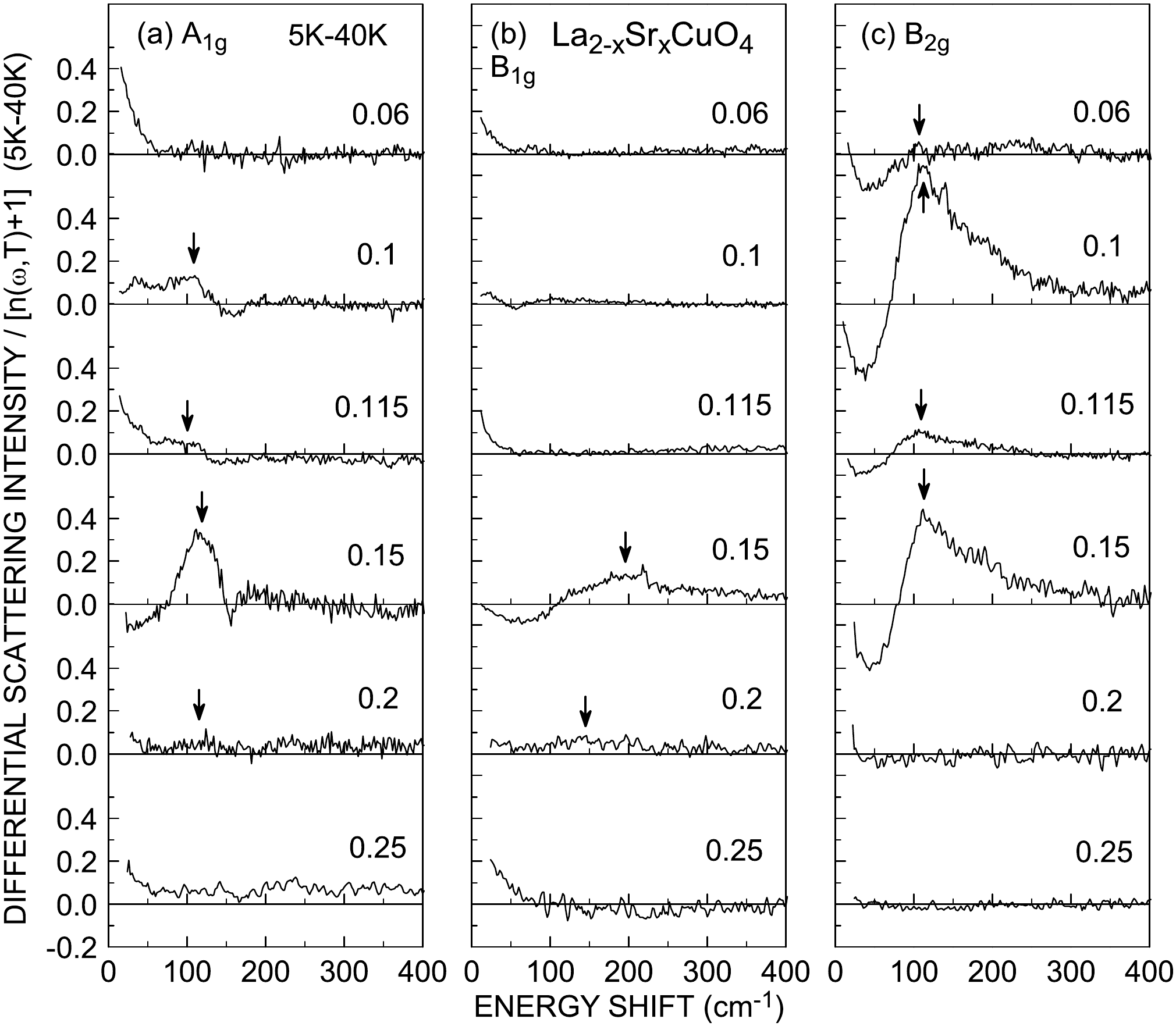}
\caption{
Differential spectra between 5 K and 40 K.  
The superconducting gap energy is indicated by the arrow.  
}
\end{center}
\end{figure*}
Figure 19 shows the differential spectra between 5 K and 40 K.  
The superconducting pair-breaking peaks are shown by the arrows.  
The gap energies (pair-breaking peak energies) are shown in Fig. 17.  
The $B_{\rm 1g}$ and $B_{\rm 2g}$ gap energies are consistent with the reported results 
\cite{Sugai,Sugai2,Sugai3,Muschler,Sugai4}.  
The $B_{\rm 2g}$ gap energies are located between the $B_1$ and $B_3$ peak energies at 
$x\le 0.15$.  
It should be noted that the $A_{\rm 1g}$ and $B_{\rm 2g}$ gap energies are independent of 
the $T_{\rm c}$.  
The $B_{\rm 1g}$ gap energy decreases with decreasing $T_{\rm c}$ at $x\ge 0.15$.  
The coupling between electrons and phonons have been observed in many experiments.  
For example, tunnel spectroscopy observed the coupling between the gap structure and 
phonons \cite{Shim}.  

The $B_{\rm 1g}$ superconducting gap at $x = 0.15$ closes above $T_{\rm c}$.  
It is different from ARPES stating that the 
pseudogap near $(\pi, 0)$ remains till $T^*=150$ K \cite{Shi,Yoshida2}.  
The $B_{\rm 2g}$ superconducting pair breaking peak appears in the polaronic 
states.  
The SDW/CDW gap and the electron-phonon coupled peaks are the fine structure of the 
Fermi arc.  
ARPES did not detect the SDW/CDW gap.  
The different results may come from the higher resolution 0.7 meV 
and the longer penetration depth 0.1 $\mu$m of light in Raman scattering 
than $15- 20$ meV and $\sim 5$ \r{A} of the electron escape depth in 
ARPES \cite{Yoshida,Shi,Yoshida2}.
The electron escape depth is shorter than the lattice constant along $c$, 13.1 \r{A}.

\section{\label{sec:pairing}Superconducting pairing model}
\subsection{\label{subsec:dislocation}Pairing at the edge dislocation of the stripe}
\begin{figure}
\begin{center}
\includegraphics[trim=0mm 0mm 0mm 0mm, width=7cm]{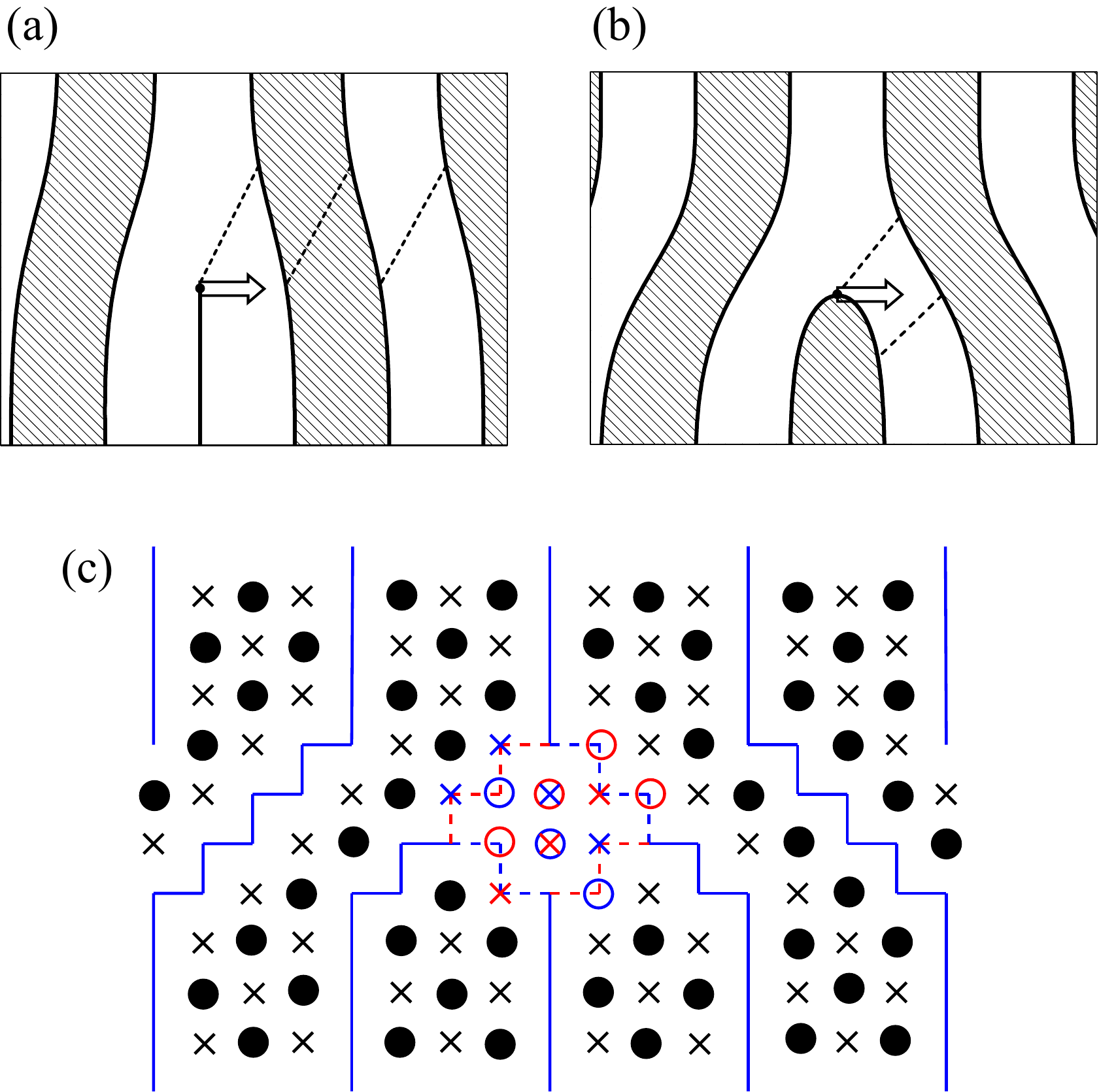}
\caption{(color online) 
Edge dislocation of (a) a single half charge stripe and (b) a looped charge 
stripe.  
The hatched and white areas have opposite spin alignment.  
(b) is more stable than (a), because the both sides of the charge 
stripe have opposite spin arrangement.  
The dashed lines show the change of the charge stripes.  
The arrow shows the Burgers vector for the movement of the edge dislocation.  
(c) Movement of the edge dislocation from the blue to the red dashed lines.  
Spins of blue open circles and christcrosses change into red ones.  
Circle: up spin, christcross: down spin, and line: charge stripe.
\label{fig:dislocation}
}
\end{center}
\end{figure}
Why does the electronic scattering show only $k\!\perp$ stripe excitations?  
In other words, why is the hole hopping restricted in the 
perpendicular direction to a stripe?  
It is reminiscent of the sliding of an edge dislocation in the Burgers 
vector direction \cite{Zaanen,Zaanen2}.  
It is well-known that ductility of metal is induced by edge 
dislocations and screw dislocations \cite{Kleinert}.  
In two-dimensional layer only edge dislocations work.  
The edge dislocation easily slides in the perpendicular direction to the 
inserted stripe.  

Figure 20(a) shows a single edge dislocation and (b) a looped edge dislocation.  
The hatched and white areas are oppositely spin ordered stripes.  
The boundaries are charge stripes.  
The open arrows are Burgers vectors.  
The Burgers vector is a vector that represents the direction and magnitude 
of the lattice distortion in a crystal.  
The edge dislocation of the looped charge stripe in Fig. 20(b) has lower 
energy than the single half change stripe in Fig. 20(a), because stable 
spins are antiparallel on both sides of the charge stripe \cite{Zaanen}.  
The dashed lines show displacements of charge stripes for the 
sliding of the edge dislocation.  
The edge dislocation easily slides perpendicularly to the stripe only 
by the local atomic displacement.  
While, the motion in the stripe direction is difficult, because new charges 
have to move from far sites.  
Charge transfer is united with the sliding of the edge dislocation.  
Other charges are localized, because the $k\parallel$ stripe excitations 
do not appear in the $B_{\rm 2g}$ Raman spectra.  
Most of the stripe structure is static except for the edges.  
The charge hopping only at the edge dislocation keeping other charges localized 
may cause the very short mean free path called ``bad metal'' \cite{Emery1,Emery2}.  
The mean free path $l$ is so short $k_{\rm F}l\approx 0.1$ that violates the 
Mott limit for the metallic transport \cite{Ando}.  
The $T$-linear resistivity \cite{Ito,Ando,Sugairesistivity} at the optimum 
doping may be induced by the present charge transfer mechanism.  

Figure 20(c) shows a model for the movement of an edge dislocation.  
The dislocation moves from the initial state (blue) to the final 
state (red) in the direction of the Burgers vector.  
The circle (christcross) indicates up (down) spin.  
The up (down) spin number changes from 3 (4) to 4 (3).  
Thus the movement of the dislocation induces the magnetic 
excitation.  
Two charged Cu atomic sites on the looped edge dislocation shift to 
right and three charged sites on the right neighbor stripe shift to left.  
The charge density on the charge stripe is a half hole per Cu site 
at $x\le 1/8$.  
Then one hole moves to right and one or two holes move 
to left.  
The distance between two holes moving to the opposite directions are 
of the order of the inter-charge stripe distance.

\begin{figure}
\begin{center}
\includegraphics[trim=0mm 0mm 0mm 0mm, width=8cm]{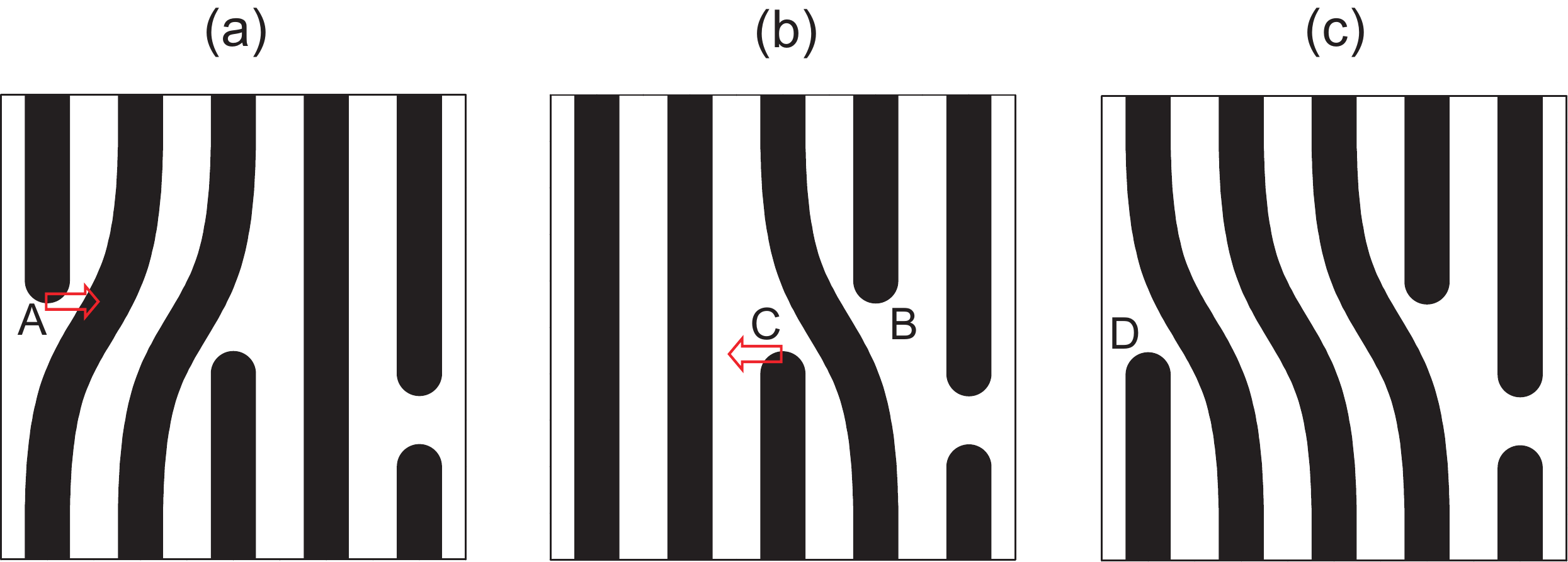}
\caption{(color online)
The edge dislocation A in (a) moves to B in (b).  
Then the edge dislocation C moves to D in (c).  
The black and white stripes denote the two different antiferromagnetic 
spin alignment and the boundary is the charge stripe.
}
\end{center}
\end{figure}
Edge dislocations in metal easily move far away.  
We suppose the same is true in the stripes of LSCO.  
Figure 21 shows a snapshot of edge dislocations.  
An edge dislocation A in Fig. 21(a) moves to B in Fig. 21(b).  
Then an edge dislocation C in Fig. 21(b) moves to D in Fig. 21(c).  
Many parts of the parallel stripe structure do not change.  
It is the reason that quasi-elastic neutron scattering can detect the stripe 
structure.  

The $B_{\rm 1g}$ phonon peak at 216 cm$^{-1}$ ($x=0$) separates into the original sharp 
peak at 218 cm$^{-1}$ and the satellite broad peak at 244 cm$^{-1}$ ($x=0.06$) in Fig. 13(b).  
The satellite peak is also infrared active \cite{Padilla}.  
The regular stripe structure has the inversion symmetry, but the edge dislocation in Fig. 20(c) 
has not the inversion symmetry.  
The Raman and infrared activities are exclusive in the crystal structure with the inversion 
symmetry.  
The phonon at the regular stripe structure is Raman active, while the localized 
phonon at edge dislocations is both Raman and infrared active.  
Therefore the original sharp peak is derived from the phonon mode at the regular stripes, 
and the satellite broad peak is derived from edge dislocations.  
The relative intensity of the satellite peak increases, as the carrier density increases.  
It is consistent with the increase of the dislocation density with the increase of carrier 
density.  
Near the optimum doping the pseudogap disappears and the $B_{\rm 1g}$ scattering intensity 
becomes stronger than the $B_{\rm 2g}$ intensity as argued in Section~\ref{subsec:eleram}.  
In the overdoped phase the dislocation density strongly increases and the movement 
disturbs the stripe structure.  
The electronic states change into the normal metal at $x\approx 0.28$.  
At the same time the stripe component disappears in neutron scattering \cite{Wakimoto2004}.  .

\subsection{\label{subsec:coherence}Coherence length}
\begin{figure}
\begin{center}
\includegraphics[trim=0mm 0mm 0mm 0mm, width=7cm]{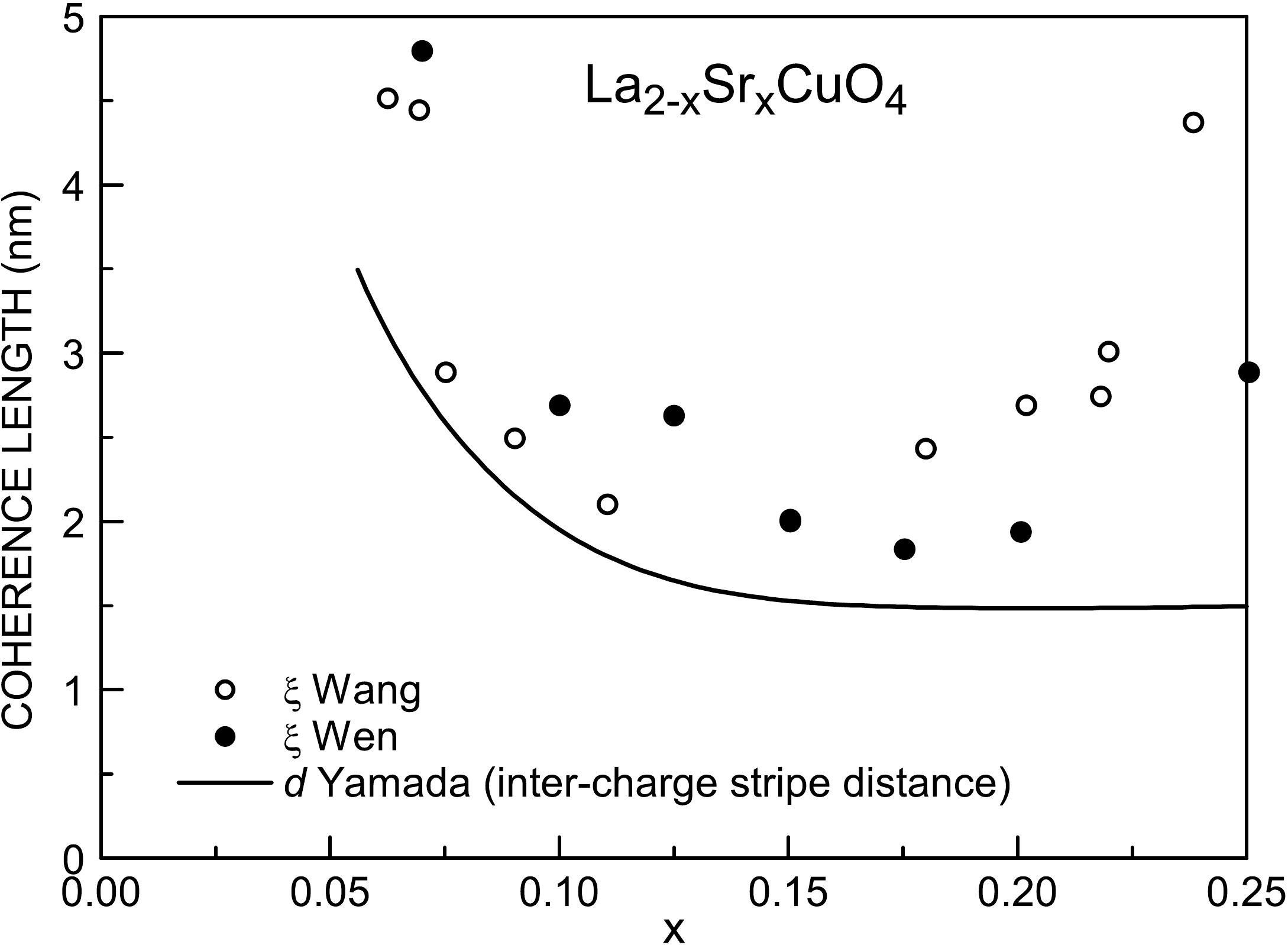}
\caption{
Superconducting coherence length $\xi$ \cite{Wang,Wen} and the inter-charge 
stripe distance $d$ \cite{Yamada}.
}
\end{center}
\end{figure}
The superconducting coherence length $\xi$ is the size of superconducting pairs.  
It is known that the common coherence length $\xi=1.5$ nm of many hole-doped high 
temperature superconductors is exceptionally short \cite{Wang,Wen,Gao,Wang2003}.  
It is in the crossover region of the BCS-BEC diagram \cite{Melo,Tsuchiya}.  
Figure 22 shows the carrier density dependence of the coherence 
length \cite{Wang,Wen} and the inter-charge stripe distance \cite{Yamada}.  
Both are surprisingly close at $x\le 0.2$.  
It supports the model that two holes at the looped edge dislocation 
form a pair.  
The increase of the $\xi$ at $x>0.2$ may be related to the increase of the edge 
dislocation density and the stripe structure are changing into the normal metallic state.  
The coherence length is only twice the inter-charge distance, 
$\sqrt{a^2/x}$, where $a$ is the Cu-Cu distance 
on the assumption that all doped carriers form pairs.  
If we take into account the instantaneous picture that many carriers 
except for edges are localized, the overlap of pairs is much reduced.  
In the weak coupling BCS regime the Fermi surface is crucial for 
the stability of the superconducting state, but in the strong BEC region 
the Fermi surface is not important.  
As a result the high temperature superconducting state appears 
in spite of a pseudogap and a SDW/CDW gap.

\section{\label{sec:pseudogap}Pseudogap}
\begin{figure*}
\begin{center}
\includegraphics[trim=0mm 0mm 0mm 0mm, width=17cm]{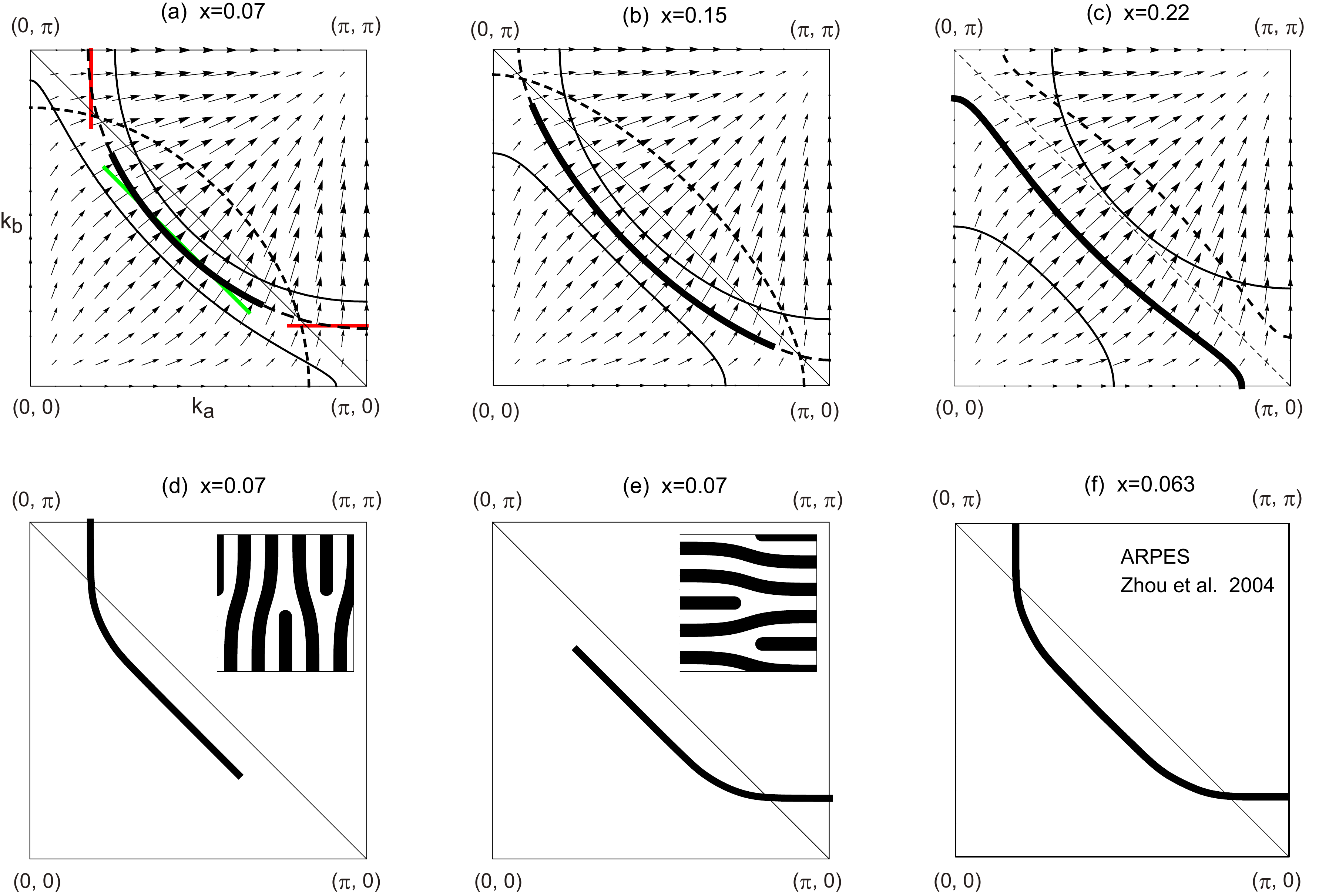}
\caption{(color online)
The Fermi surface and the group velocity (arrow) for the dispersion of 
Eq.~(\ref{eq:tightbinding}) at (a) $x=0.07$, (b) $x=0.15$, and (c) $x=0.22$.
The thick line is the Fermi arc of polaron states and the extending dashed line is the pseudogap.  
The red lines near $(0,\pi)$ and $(\pi,0)$ are flat Fermi surfaces for the 
charge transfer in the $a$ and $b$ direction, respectively.  
The green line is the Fermi surface perpendicular to the $(\pi,\pi)$ phonons, because the 
electronic states near $(\pi/2,\pi/2)$ strongly interact with the $(\pi,\pi)$ phonons.  
The two thin lines indicate the positions of $E=\pm 1000$ cm$^{-1}$ in (a) and (b) and 
$E=\pm 2500$ cm$^{-1}$ in (c).  
The dashed thin line shows the shadow Fermi surface.  
The line connecting $(\pi,0)$ and $(0,\pi)$ in (a) and (b) is the Brillouin zone boundary of the 
orthorhombic structure and the antiferromagnetic structure.  
The dashed line connecting $(\pi,0)$ and $(0,\pi)$ in (c) is the antiferromagnetic 
Brillouin zone.  
The crystal structure at $x=0.07$ and 0.15 is orthorhombic, while that at $x=0.22$ is 
tetragonal at 10 K.  
(d) and (e) show the Fermi surfaces for the stripes parallel to $b$ and $a$, respectively.  
The insets show the stripe structure.  
The black and white areas have the different spin arrangement.  
the boundaries between the black and white areas are the charge stripes.  
(f) shows the Fermi surface at $x=0.063$ obtained by Zhou {\it et al}. in ARPES \cite{Zhou2004}.
}
\end{center}
\end{figure*}
The pseudogap was first found in nuclear magnetic resonance (NMR) \cite{Yasuoka}.  
The pseudogap is observed in NMR \cite{Yasuoka1997}, resistivity \cite{Ito,Ando2004}, 
magnetic susceptibility \cite{Nakano}, infrared 
spectroscopy \cite{Lee2005,Hwang}, polarized neutron diffraction \cite{Fauque,Li2008,Mook2008}, 
tunnel spectroscopy \cite{Kohsaka2008}, 
Polar Kerr-effect \cite{Xia2008}, Nernst effect \cite{Daou2010}, ARPES, and many other experiments 
\cite{Timusk1999,Norman2005}.  
Many pseudogap models including the preformed superconducting pairs 
\cite{Anderson,YangKY,LeBlanc,Emery1,Emery3,Granath}, antiferromagnetic correlation 
\cite{Kamimura,Prelovsek}, and a density wave  \cite{Chakravarty,Sedrakyan}
were proposed.  
ARPES reported that the pseudogap opens at the anti-nodal region near $(0,\pi)$ and $(\pi, 0)$ below 
$T^*$ on the $d$ wave superconducting gap curve (one-gap model) \cite{Norman1998,Yang,Kanigel}.  
Recent ARPES, however, reported that the pseudogap energy is much higher than the 
extrapolated $d$ wave superconducting gap energy (two-gap model) in 
LSCO \cite{Terashima,Yoshida2}, 
Bi$_{2-y}$Pb$_y$Sr$_{2-x}$La$_x$CuO$_6$ (Bi2201) 
\cite{Kondo,Kondo2009,Hashimoto,Hashimoto2010}, 
and Bi$_2$Sr$_2$Ca$_{1-x}$Y$_x$Cu$_2$O$_8$ (Bi2212) \cite{Kondo2009,Tanaka,Lee}.
The energy is about 80 meV (640 cm$^{-1}$) at the insulator-metal transition point in 
LSCO \cite{Yoshida} and Bi2212 \cite{Tanaka}.
Hashimoto {\it et al}. \cite{Hashimoto2010} observed the particle-hole symmetry breaking 
in Bi2201, indicating that the pseudogap is distinct from the preformed superconducting gap.  

We propose a new model based on our finding that the charge transfer is restricted only in 
the direction perpendicular to the stripe.  
Figure 23(a) shows the Fermi surface (thick solid line and the extending dashed line) and 
the group velocity (arrow) for the energy dispersion of Eq.~(\ref{eq:tightbinding}) 
\cite{Yoshida} at $x=0.07$.  
The velocity is perpendicular to the Fermi surface.  
A quarter of the tetragonal Brillouin zone is shown.  
If the stripe is parallel to the $b$ axis, the allowed charge hopping direction is $a$.  
One-dimensional conductor has a flat Fermi surface perpendicular to the conducting direction.  
The velocity of the Fermi surface near $(0,\pi)$ is parallel to the allowed charge transfer 
direction, but that near $(\pi,0)$ is orthogonal to the allowed direction.  
Therefore the electronic transition across the Fermi surface near $(\pi,0)$ is suppressed.  
It is observed as the pseudogap.  
The $B_{\rm 1g}$ electronic scattering spectra becomes the same as the $B_{\rm 2g}$ spectra 
above 2000 cm$^{-1}$ in the underdoped phase as discussed in Section~\ref{subsec:high}.  
It was understood that the isotropy in $k$ space for the electronic transition increases 
as the energy shift increases and the transition becomes completely isotropic above 2000 
cm$^{-1}$ in the underdoped phase.  
The positions of $E=\pm 1000$ cm$^{-1}$ are shown by two thin solid curves in Fig. 23(a), 
although the isotropy in $k$ space indicates that the momentum is not a good quantum number.    
The curve on the $(0,0)$ side is $E=-1000$ cm$^{-1}$ and that on the $(\pi,\pi)$ side is 
$E=1000$ cm$^{-1}$.  
The transition within these two curves is anisotropic and shows the pseudogap near $(\pi,0)$.  
The short $k$ transition corresponds to the long-range transfer more than ten 
times the lattice constant in the real space.  
The pseudogap closes for the transition from the outer side including $(0,0)$ to the outer 
side including $(\pi,\pi)$.  
The stripe direction is fluctuating in the $a$ or $b$ direction.  
For the stripe parallel to $a$, the Fermi surface near $(0,\pi)$ has a pseudogap.  
Figure 23(b) shows the Fermi surface at the optimum doping $x=0.15$.  
The pseudogap is plotted so that the velocity on the Fermi surface has the same range of 
gradient as in the pseudogap at $x=0.06$.  
The pseudogap decreases, because the position of the Fermi surface in $k$ space changes.  
The thin solid curves indicate the $E=\pm 1000$ cm$^{-1}$ positions.  
Figure 23(c) shows the Fermi surface in the overdoped phase at $x=0.22$.  
The velocity is not perpendicular to the $a$ axis on the almost whole Fermi surface 
except for the very small spot on the $(0,0)-(0,\pi)$ line.  
Therefore the pseudogap does not appear.  
Thus the carrier density dependence of the pseudogap is naturally explained in the 
restricted charge transfer direction to $a$ or $b$.  
The boundary of the anisotropic-isotropic excitations is $4000-5000$ cm$^{-1}$ in the 
overdoped phase.  
The thin solid curves indicate the $E=\pm 2500$ cm$^{-1}$ positions.  

A one-dimensional conductor has a flat Fermi surface.  
The tight binding Fermi surface for the stripes parallel to $b$ is rounded near $(0,\pi)$ 
at $x=0.07$ in Fig. 23(a).  
If the Fermi surface near $(0,\pi)$ becomes flat and perpendicular to the $a$ axis as shown by 
the red line, the charge transfer increases and the kinetic energy decreases, 
because the group velocity is perpendicular to the Fermi surface.  
The Fermi surface for the stripes parallel to $b$ is shown in Fig 23(d).  
The flat region near $(\pi/2,\pi/2)$ comes from a different mechanism as discussed later.  
In the same way the Fermi surface near $(\pi,0)$ becomes flat in Fig. 23(e) to decrease the 
kinetic energy for the stripes parallel to the $a$ axis.  
In the crystal of mixed stripe directions the observed Fermi surface is the average of 
Fig. 23(d) and (e).  
In fact the flat Fermi surface was observed near $(0,\pi)$ and $(\pi,0)$ at $x=0.063$ and 1/8 
in ARPES \cite{Zhou1999,Zhou2004,Valla2006}.  
Figure 23(f) shows the Fermi surface at $x=0.063$ obtained by Zhou {\it at al}. \cite{Zhou2004}.  
The one-dimensional charge transfer along the stripe was considered in ARPES \cite{Valla2006}, 
but the present experiment revealed that it is perpendicular to the stripe.  
The Fermi surface measured by ARPES has four-fold rotational symmetry, because 
the stripe direction is fluctuating in space and time.  
But the Fermi surface of the stripe phase has no four-fold rotational symmetry 
as shown in Fig. 23(d) and (e).  
The four-fold rotational symmetry breaking was observed in tunnel spectroscopy \cite{Kohsaka2008} 
and Nernst effect \cite{Daou2010}.  

Another model to break the rotational symmetry is the $d$-wave Pomeranchuk 
instability \cite{Yamase,Halboth}.  
Yamase and Zhyher \cite{Yamase2011} calculated the Raman susceptibility near the 
$d$-wave Pomeranchuk instability.  
The $d$-wave Pomeranchuk instability couples to the $B_{\rm 1g}$ electronic and 
phononic excitations.  
A central peak emerges at the energy shift zero for each of the electronic and 
phononic $B_{\rm 1g}$ spectra, as temperature decreases in the carrier density 
below the critical value $(x\le x_{\rm c})$.  
The central peaks change into two low-energy peaks for the electronic and phononic 
channels at $x>x_{\rm c}$.  
The soft mode energies increase with broadening, as the carrier density increases.  
The $B_{\rm 1g}$ spectra in Fig. 13(b) have not such a central peak nor the 
low-energy peak whose energy increases with increasing the carrier density.  
The $B_{\rm 1g}$ phonon of 218 cm$^{-1}$ has the satellite peak on the high 
energy side.  It is the opposite side of the prediction from the Pomeranchuk model.  
Therefore the present Raman scattering experiment gives a negative result for 
the Pomeranchuk instability.  

The electron-phonon coupled hump 
below 180 cm$^{-1}$ and the magnetic hump from 1000 cm$^{-1}$ to 3500 cm$^{-1}$ 
are strongly enhanced near the insulator-metal transition at low temperatures 
in the $B_{\rm 2g}$ spectra of Fig. 4 and 13.  
The electronic states near $(\pi/2,\pi/2)$ strongly interact with the B$_2$, 
B$_3$, and B$_4$ phonons as discussed in Section~\ref{subsec:low}.  
The B$_2$, B$_4$ modes are the $(\pi, \pi)$ phonon mode.  
The B$_3$ mode cannot be determined whether it is the $(0,0)$ mode or 
$(\pi,\pi)$ mode, because the dispersion is flat \cite{Rietschel1989}.  
If one assumes this mode to be the $(\pi,\pi)$ mode, all the modes are the zone 
boundary modes.  
The momentum $(\pi,\pi)$ is the reciprocal lattice vector to form the orthorhombic 
structure from the tetragonal structure and also the antiferromagnetic structure.  
Usually the structural phase transition is induced by the softening of a single 
phonon.  
It is the A$_1$ phonon in Fig. 14 \cite{Chaplot,Boni}.  
In the present case the electronic states strongly interact with many phonons 
with the momentum producing the lower symmetry structure.  
It is rather anomalous.  
If the electronic states with the velocity parallel to $(\pi,\pi)$ is preferable to 
stabilize the system through the electron-many phonon interactions, the Fermi 
surface changes to increase the part in which the velocity is parallel to $(\pi,\pi)$.  
It is shown by the green line in Fig. 23(a).  
The electron-phonon coupled hump below 180 cm$^{-1}$ is largest at $x=0.06$ in 
Fig. 13(c).  
At the almost same carrier density at $x=0.063$ Zhou {\it et al}. \cite{Zhou2004} 
observed the flat Fermi surface perpendicular to $(\pi,\pi)$ at the large area around 
$(\pi/2,\pi/2)$ in ARPES as shown in Fig. 23(f).  
It is supposed that the orthorhombic structure is stabilized by the dynamic coupling 
between the electronic states near $(\pi/2,\pi/2)$ and many $(\pi,\pi)$ phonons.  
It is, however, not determined whether the phonon wave vector is exactly $(\pi,\pi)$ 
or a little shorter to nest the Fermi surfaces near $(\pi/2,\pi/2)$ and 
$(-\pi/2,-\pi/2)$, because the phonon dispersions near $(\pi,\pi)$ are nearly flat.  
In the latter case the phonons work to increase the nesting susceptibility.  

The thin dashed line in Fig. 23(a), (b) and (c) is the shadow Fermi surface which 
is the $(\pi,\pi)$ shifted primary Fermi surface.  
It is the folded Fermi surface in the Brillouin zone of the 
orthorhombic structure and also the antiferromagnetic structure.  
The crystal structure is orthorhombic at $x=0.07$ and 0.15 and tetragonal at 
$x=0.22$.  
The shadow Fermi surface is observed in ARPES of Bi2212 \cite{Mans,Meng}, Bi2201 
\cite{Nakayama}, and LSCO \cite{Zhou2007,Chang}.  
The Fermi pocket is observed in Bi2212 \cite{Meng}.  
The shadow Fermi surface in LSCO is observed in the underdoped phase, but not in the 
overdoped phase \cite{Zhou2007,Chang}.  
The magnetic hump from 1000 cm$^{-1}$ to 3500 cm$^{-1}$ is small at $x\approx 1/8$ 
in Fig. 4(c), while the shadow Fermi surface is observed \cite{Chang}.  
Therefore the shadow Fermi surface is induced by the lattice effect in agreement 
with Mans {\it et al} \cite{Mans}.  

In the underdoped insulating phase ($x<0.055$) the stripe direction changes into 
the diagonal direction \cite{Wakimoto}.  
However, the $B_{\rm 1g}$ and $B_{\rm 2g}$ spectra at $x=0.035$ in Fig. 4 and 13 
does not change qualitatively from the spectra in the metallic phase.  
Seibold and Lorenzana \cite{Seibold2} calculated the $k||$ and $k\!\perp$ stripe 
dispersions for magnetic excitations in the site-centered and bond-centered stripe 
structure at $x=0.05$.  
It is difficult to assign the Raman data to the dispersions, because the number 
of dispersion segments is too many.  
In the calculation the intensity of the $k\!\perp$ stripe magnetic susceptibility 
is weak at the intermediate energy range \cite{Seibold2}.  
The $B_{\rm 2g}$ spectra in Fig. 4(c) do not show a decrease at the middle of the 
hump from 1000 to 3500 cm$^{-1}$.  
The pseudogap is observed at $(0,\pi)$ and $(\pi,0)$ in the extrapolated shape 
from the metallic phase in ARPES \cite{Yoshida}.
In the diagonal stripe parallel to $(\pi,\pi)$ the Burgers vector is parallel to 
$(-\pi,\pi)$.  
The pseudogap opens near $(\pi/2,\pi/2)$, if our mechanism of the pseudogap is 
applied to the insulating phase.  
But the experimental results are different.  
Therefore it is supposed that the charge transfer is large in the nearest neighbor 
direction $a$ or $b$.  
The resistivity of LSCO with $x=0.03$ decreases on decreasing temperature from high 
temperature to 70 K in the same way as the metallic phase and then the resistivity 
increases below 70 K \cite{Ando,Sugairesistivity}.  
It may be explained as follows.  
The effect of the different directions between the charge transfer and the Burgers 
vector is relaxed by the thermal excitation at high temperatures, 
but the difference becomes crucial at low temperatures and the resistivity increases.  
La$_2$NiO$_{4+\delta}$ with the diagonal stripe structure \cite{Tranquada} is an 
insulator, too.

The high energy excitations comes from the short range electronic excitations.  
The excitations in short distance is very complicated by the rearrangement 
of spins and charges in the moving looped edge stripe in Fig. 20(c).  
It may be the origin of the isotropic energy state in $k$ space.  
The pseudogap energy is 2000 cm$^{-1}$, if it is estimated from the split of the 
$B_{\rm 1g}$ spectra from the $B_{\rm 2g}$ spectra in Fig. 6 and 7.
This energy is independent of the carrier density and temperature in the underdoped phase.  
The pseudogap energy observed by ARPES is about 80 meV (640 cm$^{-1}$) at the 
insulator-metal transition \cite{Tanaka,Yoshida2}.  
Many ARPES experiments reported that the gap energy depends on the carrier density 
and the gap closes at $T^*$ \cite{Norman1998,Shi,Yoshida2,Kondo,Kondo2009,Hashimoto2010}.  
However, ARPES also reported the example that the pseudogap survives far above 
$T^*$ \cite{Kordyuk}.  
The large energy difference comes from the fact that (1) 2000 cm$^{-1}$ is 
the highest energy of the different $B_{\rm 1g}$ and $B_{\rm 2g}$ spectra and not the 
direct gap energy and (2) Raman scattering observes the energy from the valence band 
to the conduction band, while ARPES observes the energy from the valence band to the 
chemical potential.  

\section{\label{sec:discussions}Discussions}
\begin{figure}
\begin{center}
\includegraphics[trim=0mm 0mm 0mm 0mm, width=7cm]{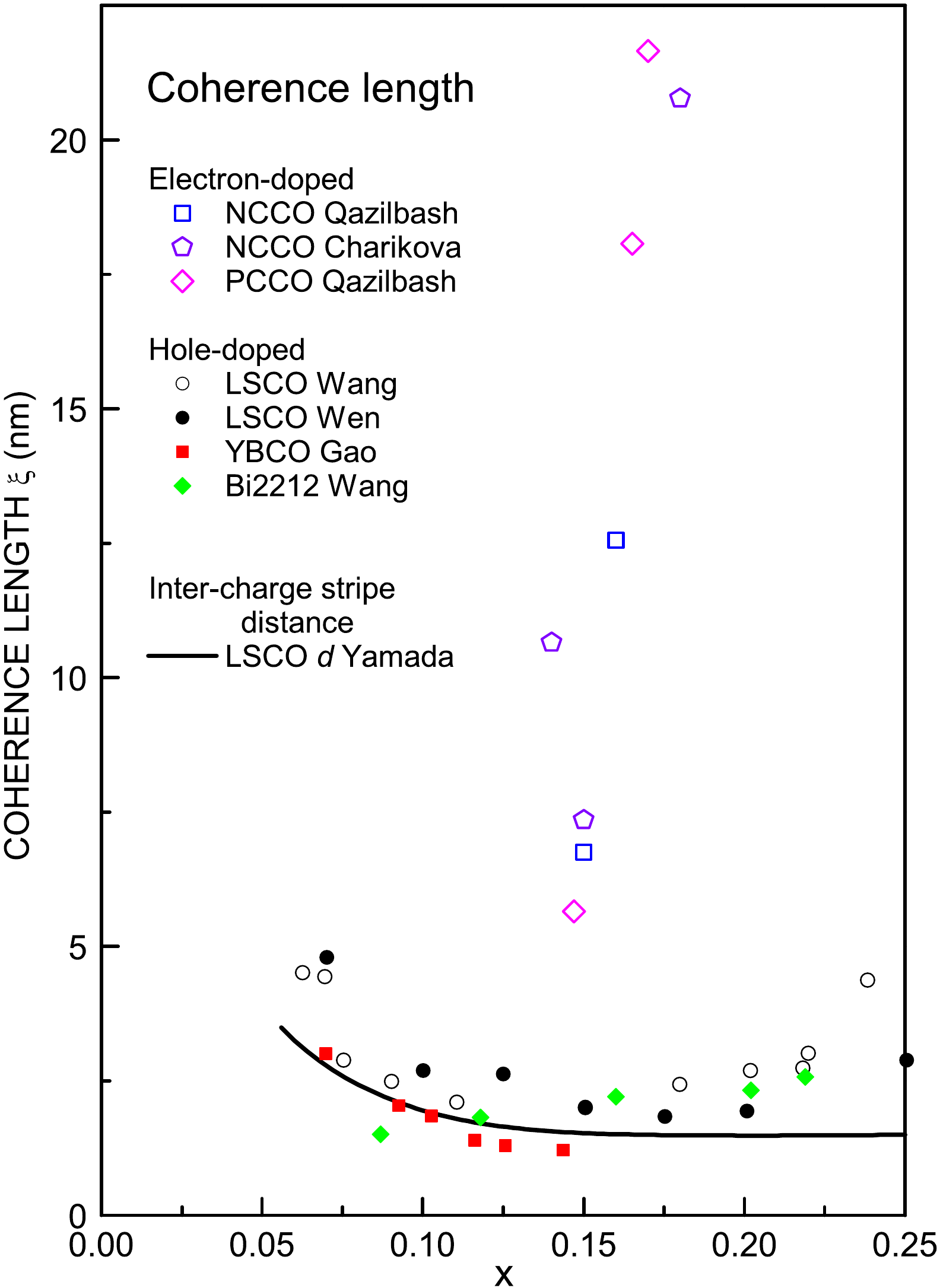}
\caption{
Superconducting coherence length $\xi$ in hole-doped LSCO \cite{Wang,Wen}, YBCO \cite{Gao}, 
Bi2212 \cite{Wang2003}, and electron-doped NCCO \cite{Qazilbash,Charikova} and PCCO \cite{Qazilbash}.  
The solid line is the inter-charge stripe distance $d$ \cite{Yamada}.
}
\end{center}
\end{figure}
The large difference between the hole-doped cuprate superconductors and the electron-doped 
cuprates is the existence or absence of the stripe structure.  
Neutron scattering disclosed that the magnetic scattering spot is always commensurate 
$(\pi,\pi)$ in Nd$_{2-x}$Ce$_x$CuO$_4$ (NCCO), suggesting that the stripe is absent in 
electron-doped cuprates \cite{YamadaNCCO,MotoyamaNCCO}.  
The $B_{\rm 1g}$ two-magnon peak softens on increasing the carrier density in the 
hole-doped cuprates as shown in Fig. 4(b) \cite{Sugai}.  
However, the softening of the two-magnon peak is not observed in electron-doped cuprate 
superconductors \cite{SugaiNCCO,Tomeno,Onose,SugaiLSCNCC}.  
The two-magnon peak energy does not shift in the insulating phase of NCCO, 
even if carriers are doped \cite{SugaiLSCNCC}.  
In the metallic phase the two-magnon peak disappears and the spectra shifts to much 
higher energy than the original two-magnon peak energy.  
Therefore the softening of the $B_{\rm 1g}$ two-magnon peak is not a common property in 
a doped antiferromagnet, but the property of the $k||$ stripe magnetic excitations.  
The hump from 1000 to 3500 cm$^{-1}$ observed in the $B_{\rm 1g}$ and $B_{\rm 2g}$ spectra 
in LSCO does not appear in electron doped cuprate superconductors.  
It is also the characteristic property of the stripe structure.  

Our finding that only $k\!\perp$ stripe excitations are included in the $B_{\rm 2g}$ 
spectra indicates that the charge transfer is united with the sliding motion of the edge 
dislocation which moves perpendicularly to the stripe.  
Figure 24 shows the coherence length in hole-doped cuprates and electron-doped cuprates.  
The inter-charge stripe distance of LSCO \cite{Yamada} is also shown.  
The carrier density dependence of the coherence length is almost perfectly follows 
the inter-charge stripe distance not only in LSCO but also in YBCO and Bi2212 
\cite{Wang,Wen,Gao,Wang2003}.  
On the other hand the coherence lengths of electron-doped cuprate superconductors NCCO 
and Pr$_{2-x}$Ce$_x$CuO$_4$ (PCCO) are much longer \cite{Qazilbash,Charikova}. 
It clearly indicates that the pairing is formed between charge stripes.  
The moving carriers are only at the looped edge dislocations.  
Therefore the Cooper pairs are formed at the looped edge dislocations.  

The paired charges moving with the edge dislocation is like a bi-polaron 
\cite{Alexandrov,Alexandrov2000}.  
The binding energy is, however, related to not only the electron-phonon interaction but also 
the stripe formation energy including the electron, spin and charge interactions.  
The phonons localized at the edge dislocation may not be the bulk phonons.  
The strong electron-phonon interactions are observed in the $B_{\rm 2g}$ 
channel.  
The existence of the phonon contribution is known from the isotope effect of the 
penetration length \cite{Khasanov}, although the isotope effect of the $T_{\rm c}$ is 
small at the optimum doping \cite{Pringle}. 
The contribution of phonons can introduce a retardation effect to the pairing so 
that the instantaneous Coulomb repulsion is avoided \cite{Yonemitsu,Tam,She}.

\section{\label{sec:conclusion}Conclusion}
Utilizing the different Raman selection rule between two-magnon scattering and 
electronic scattering, the $k||$ and $k\!\perp$ stripe magnetic excitations are 
separately detected in the nematic fluctuating spin-charge stripes.  
The electronic scattering has only $k\!\perp$ stripe excitations, indicating that 
the charge hopping is restricted to the direction perpendicular to the stripe.  
It is the same as the sliding of an edge dislocation in the Burgers vector direction 
which is perpendicular to the stripe.  
Consequently holes at the edge dislocations transfer together with the sliding of 
the edge dislocations.  
Other holes are localized, because the $k||$ stripe excitations are not observed 
in the electronic scattering.  
The looped edge dislocation which is made of bridged two charge stripes has lower 
energy than the single edge dislocation.  
The superconducting coherence length is surprisingly close to the inter-charge stripe 
distance at $x\le 0.2$.  
The coherence length is intermediate between the BCS and the BEC superconductors.  
Therefore it is concluded that the superconducting pairs are formed at the moving 
looped edge dislocations.  
The restricted charge transfer perpendicularly to the stripe naturally explains the 
pseudogap formation near $(0,\pi)$ or $(\pi,0)$, depending on the stripe direction.  
The parts of the Fermi surface with the pseudogap are deformed to decrease the kinetic energy.  
The electronic states near $(\pi/2,\pi/2)$ strongly interact with the $(\pi,\pi)$ phonons 
so that the Fermi arc is composed of polarons.

\end{document}